\documentclass[a4paper,11pt]{article}
\pdfoutput=1 
\usepackage{jheppub}
\usepackage{graphicx}
\usepackage{amsfonts}
\usepackage{hhline}
\usepackage{multirow}
\usepackage{array}
\usepackage{dsfont}
\usepackage{amssymb}
\usepackage{maybemath}
\usepackage{float}
\usepackage[lofdepth,lotdepth]{subfig}
\usepackage[countmax]{subfloat}
\usepackage{cancel}
\def\beq{\begin{equation}}
\def\eeq{\end{equation}}
\def\bea{\begin{eqnarray}}
\def\eea{\end{eqnarray}}

\def\roughly#1{\mathrel{\raise.3ex\hbox
{$#1$\kern-.75em\lower1ex\hbox{$\sim$}}}}

\newcommand*\xbar[1]{%
  \hbox{%
    \vbox{%
      \hrule height 0.5pt % The actual bar
      \kern0.3ex%         % Distance between bar and symbol
      \hbox{%
        \kern-0.1em%      % Shortening on the left side
        \ensuremath{#1}%
        \kern-0.1em%      % Shortening on the right side
      }%
    }%
  }%
} 

\usepackage[normalem]{ulem}
\usepackage{color}
\definecolor{BrickRed}{cmyk}{0,0.89,0.94,0.28}
\definecolor{DarkGreen}{cmyk}{1,0,1,0.5}
\definecolor{Blue}{cmyk}{1,1,0,0}
\definecolor{BurntOrange}{cmyk}{0,0.51,1,0}

\def\soutdl{\bgroup\markoverwith{\textcolor{BrickRed}{\rule[0.5ex]{2pt}{0.4pt}}}\ULon}
\def\soutps{\bgroup\markoverwith{\textcolor{DarkGreen}{\rule[0.5ex]{2pt}{0.4pt}}}\ULon}
\def\soutkk{\bgroup\markoverwith{\textcolor{Blue}{\rule[0.5ex]{2pt}{0.4pt}}}\ULon}
\def\soutas{\bgroup\markoverwith{\textcolor{BurntOrange}{\rule[0.5ex]{2pt}{0.4pt}}}\ULon}

\title{{\boldmath Implications of a Electroweak Triplet Scalar Leptoquark on the Ultra-High Energy Neutrino Events at IceCube
}}

\author[a]{Nicolas Mileo}
\author[b]{\hspace*{-4pt}, Alejandro de la Puente}
\author[a]{and Alejandro Szynkman}

\affiliation[a]{{\it IFLP, CONICET -- Dpto. de F\'{\i}sica,
    Universidad Nacional de La Plata,\\ C.C. 67, 1900 La Plata,
    Argentina}}
\affiliation[b]{{\it Ottawa-Carleton Institute for Physics, Carleton University,
 \\ 1125 Colonel By Drive, Ottawa, Ontario K1S 5B6, Canada  }}

\emailAdd{mileo@fisica.unlp.edu.ar}
\emailAdd{apuente@physics.carleton.ca}
\emailAdd{szynkman@fisica.unlp.edu.ar}

\abstract{We study the production of scalar leptoquarks at IceCube, in particular, a particle transforming as a triplet under the weak interaction. The existence of electroweak-triplet scalars is highly motivated by models of grand unification and also within radiative seesaw models for neutrino mass generation. In our framework, we extend the Standard Model by a single colored electroweak-triplet scalar leptoquark and analyze its implications on the excess of ultra-high energy neutrino events observed by the IceCube collaboration. We consider only couplings between the leptoquark to first generation of quarks and first and second generations of leptons, and carry out a statistical analysis to determine the parameters that best describe the IceCube data as well as set $95\%$ CL upper bounds. We analyze whether this study is still consistent with most up-to-date LHC data and various low energy observables. 
}
\begin{document}
\maketitle
\flushbottom
%%%
\section{Introduction}
In this work we study the implications of a colored electroweak-triplet scalar leptoquark (LQ) on the ultra high energy (UHE) neutrino spectrum observed at IceCube, focusing particularly on the range above PeV, where a bit higher than expected event rate has been reported~\cite{Aartsen:2015zva}. The potential of the IceCube facility to probe LQ models has been exploited in many works. In ref.~\cite{Anchordoqui:2006wc}, for example, the inelasticity distribution of the events detected at IceCube are used to test LQ production; in refs.~\cite{Barger:2013pla,Dutta} electroweak-singlet scalar LQs, with different flavor structure for its couplings, are introduced to fit the neutrino flux at the PeV range. In this regard, besides the many explanations that incorporate new physics effects, other possibilities within the picture of the Standard Model (SM) have also been proposed \cite{Soni,Beacom}.\par    
 Leptoquarks (LQs) are fields that arise naturally from the unification of quarks and leptons in extensions of the SM~\cite{Pati:1973uk,Georgi:1974sy,Pati:1974yy}. In particular, unification of quarks and leptons into simple groups of SU$(5)$ requires the unification of LQs with the SM-like Higgs boson. However, one main obstacle that arises from the introduction of LQs is how they can mediate proton decay at tree level, specially in the case of LQs that violate lepton and baryon numbers, if those quantum numbers are indeed assigned. Unification schemes to accommodate very heavy LQs to avoid proton decay bounds have also been studied, in particular a scheme based on a flipped SU$(5)$ framework where SM fields are embedded into representations of a SU$(5)\times$U$(1)$ gauge group has proved successful~\cite{Barr:1981qv,DeRujula:1980qc,Derendinger:1983aj,Antoniadis:1987dx}. 
However, in view of the current experimental effort to produce particles beyond the SM, most studies have focused on two particular scalar LQ representations out of the six possible ones~\cite{Buchmuller:1986zs}, where phenomenologically light LQs are natural. These fields transform as ${\bf (3,2,1/6)}$ and ${\bf (3,2,7/6)}$ under the SM SU$(3)_{\text{c}}\times$SU$(2)_{\text{W}}\times$U$(1)_{\text{Y}}$  gauge group and have been implemented to address several hints of new physics beyond the SM, in particular the excess reported by IceCube~\cite{Dey:2015eaa} and the anomalous LHC same-sign lepton events~\cite{CMS:2014jfa} such as~\cite{Queiroz:2014pra,Allanach:2015ria}. These two weak doublets do not couple to baryon number violating operators at tree level. However, effects of higher dimensional operators can cause baryon number violation. In this regard, the authors in~\cite{Arnold:2013cva} discuss a framework where one can naturally suppress these operators. Despite the fact that the two representations mentioned above are the most frequently used, other LQ models with diquark operators have been also considered to address other reported anomalies. One recent work, for example, uses a LQ with the quantum numbers ${\bf (3,1,-1/3)}$ to address deviations on $R_{D^{*}}$, $R_{K}$, and the $(g-2)$ of the muon~\cite{Bauer:2015knc} and similarly using the electroweak doublets introduced above~\cite{Sahoo:2015fla,Sahoo:2015pzk,Sahoo:2016slm,Sahoo:2016nvx}. Another work uses this electroweak singlet LQ to explain the excess of high energy neutrino events~\cite{Barger:2013pla}. Only two other scalar LQ representations can couple to SM neutrinos and quarks and are thus relevant in the explanation of the IceCube excess, these are the ${\bf (3,1,-1/3)}$ mentioned earlier and a weak triplet ${\bf (3,3,-1/3)}$. In contrast to the former, the latter has not been probed through the UHE neutrino spectrum observed at Icecube. Both LQs couple to diquarks and can induce proton decay at tree level. The authors in~\cite{Dorsner:2009cu} discuss a scenario to suppress the diquark operators by embedding the weak triplet and singlet into a ${\bf45}_{H}$-dimensional Higgs representation of a SU$(5)$ GUT model. It is therefore plausible to consider light weak triplet and singlet LQs, with masses accessible at colliders, as a possible source of the UHE neutrino events observed at IceCube. The study of LQs has become very active; with a focus also on $R$-parity violating scenarios of supersymmetry (SUSY) which yield couplings of scalar superpartners to quarks and leptons. As far as the UHE neutrino events observed at IceCube is concerned, these can be used to constrain $R$-parity violating supersymmetric models~\cite{Carena:1998gd,Dev:2016uxj}. LQ have a rich phenomenology and for this reason we direct the reader to a recent review~\cite{Dorsner:2016wpm} for more an in depth discussion and references therein.

In this work we focus on the weak triplet since this class of particles has recently been used to mediate the generation of neutrinos masses radiatively and at three loops~\cite{Ng:2013xja,Ng:2014pqa}. The model considered by these authors also includes a heavy Majorana neutrino dark matter candidate. The work focuses solely on the phenomenology of a LQ coupling right-handed up-type quarks to the Majorana neutrino, yielding a mechanism for its relic abundance. In addition, a monotop search strategy was introduced and limits were placed on the model using current LHC data. In this work, we wish to go one step further, that is, analyze the phenomenology of the weak triplet, originally with masses set at the TeV scale, and introduce a mechanism to produce high energy neutrino events in detectors such as IceCube. Our model is very attractive since the coupling of the LQ to up- and down-type quarks is the same. It also allows us to directly connect the observations by the IceCube collaboration to the mechanism of neutrino mass generation and specific GUT scenarios where tree-level baryon number violating operators are absent. 
%For an extensive review on LQs and also on an extensive analysis of those mediating proton decay as well as the different unification schemes one may incorporate we refer the reader to~\cite{Dorsner:2012nq,Dorsner:2016wpm}

The remainder of this paper is organized as follows. In sec.~\ref{sec2} we review the model specifying those aspects related to the  UHE neutrino events at IceCube. In sec.~\ref{sec3} we probe the proposed weak triplet with the IceCube detector. We review the SM neutrino-nucleon scattering cross section and compute the respective LQ contribution in sec.~\ref{sec3.1}. We then obtain in sec.~\ref{sec3.2} the new physics contribution to the rate of events expected at IceCube and study its behaviour with respect to the LQ masses. In sec.~\ref{sec3.3} we perform an statistical analysis in order to determine the parameters that best fit the IceCube data as well as set upper limits as a function of the LQ mass. Secs.~\ref{sec3} and \ref{sec4} are dedicated to the analisis of constraints arising from the LHC experiments with the $8$ and $13$ TeV data sets, lepton flavor violation (LFV), and low energy precision measurements such as atomic parity violation. Finally, in sec.~\ref{sec6}, we compare the results obtained from the analysis of the IceCube data with the constraints derived in secs.~\ref{sec4} and \ref{sec5} and provide some concluding remarks. The Appendix gives some further details about the attenuation effects on upward-going neutrinos resulting from their passage through the Earth.
%%%%%%%%%%%%%%%%%%%%%%%%%%%%%%%%%%%%%%%%%%%%%%%%%%%%%%%%%%%%%%%%%%%%
\section{Model}
\label{sec2}
The authors in~\cite{Ng:2013xja,Ng:2014pqa} investigated a model that incorporates LQs, one triplet under the SU$(2)_{\text{W}}$ gauge interaction and a singlet. In addition, the model contains a single Majorana right-handed neutrino used to both explain the nature of dark matter and the mechanism for neutrino mass generation. Within this framework Majorana masses for the active neutrinos were made possible via a radiative process which involves a three loop diagram. In order for the mechanism to work, two representations of weak triplets were implemented: a lepton and baryon number violating LQ transforming as a ${\bf (3,3,-1/3)}$ under the SM gauge group and a weak triplet transforming as a ${\bf (3,3,2/3)}$ with no tree-level coupling to fermions. In this work we are primarily interested in the former because it couples quarks to leptons and therefore affects the neutrino-nucleon cross section, which may lead to new features in the spectrum of UHE neutrinos observed by the IceCube collaboration. 
In the following we will refer to the field transforming as a ${\bf (3,3,-1/3)}$ under the SU$(2)_{\text{W}}\times \text{U}(1)_{\text{Y}}$ SM interactions as $\chi$. Given its quantum numbers, one may choose to write $\chi$ as a $2\times 2$ matrix with the following transformation property
\begin{equation}
\label{eq2.1}
\chi\to U\chi U^{\dagger},
\end{equation}
where $U=\text{exp}(i\omega_{j}\tau_{j}/2)$ and $\tau_{j}$ is the $j$-th Pauli matrix. We then represent the weak triplet $\chi$ with the following matrix:
\begin{equation}
\label{eq2.2}
\begin{pmatrix}
\chi_2/\sqrt{2} & \chi_1 \\
\chi_3 & -\chi_2/\sqrt{2} \\
\end{pmatrix},
\end{equation}
and parametrize its interactions with left-handed quarks and leptons with the following Lagrangian:
\begin{equation}
\nonumber
\mathcal{L}_{LQ}\supset \lambda^i_{j}\left[\xbar{u}_{\,iL}\left(-\chi_1\,\nu^c_{jL}+\frac{\chi_2}{\sqrt{2}}\,e^{\,c}_{jL}\right)+\xbar{d}_{iL}\left(\frac{\chi_2}{\sqrt{2}}\,\nu^c_{jL}+\chi_3\,e^{\,c}_{jL}\right)\right]+\mathrm{h.c.}
\end{equation}
\begin{equation}
\label{eq2.3}
\qquad=\lambda^i_{j}\left[\xbar{u}_{\,i}P_R\left(-\chi_1\,\nu^c_{j}+\frac{\chi_2}{\sqrt{2}}\,e^{\,c}_j\right)+\xbar{d}_{\,i}P_R\left(\frac{\chi_2}{\sqrt{2}}\,\nu^c_{j}+\chi_3\,e^{\,c}_j\right)\right]+\mathrm{h.c.}\,,
\end{equation}
where $P_R=(1+\gamma_5)/2$, $\lambda^i_{j}$ represents the coupling between the $i$-th generation of quarks and the $j$-th generation of leptons with $i,j=1,2,3$, and $\psi^c$ denotes the conjugate field of $\psi$.\par 
The terms in the above Lagrangian are not the only ones allowed by gauge invariance. One can incorporate a quark bilinear operator coupling to the weak triplet given by
\begin{equation}
\label{eq2.4}
{\cal L}_{QQ}\supset y_{ij}\,\xbar{Q}^{\,c}_{iL}(i\tau_2\chi)Q_{jL},
\end{equation}
where the indices $i$ and $j$ run over the three quark generations and $Q_L$ denotes the quark weak doublet. The interaction in eq.~(\ref{eq2.4}) induces rapid proton decay and a symmetry needs to be imposed in order to suppress the strength of these interactions. However, as shown in~\cite{Barr:2012xb}, the above operator also induces a Planck scale suppressed dimension five operator that gives the decay modes $p\to\pi^{+}\nu$, $p\to K^{+}\nu$ and $p\to K^{+}\pi^{+}l^{-}$. In order to generate two-body nucleon decay partial rates near the present limits one would require 
\begin{equation}
\label{eq2.5}
m_{\chi}\sim(3k)^{1/4}(y\cdot Y_{5})^{1/2}10^{7}~\text{GeV},
\end{equation}
where $Y_{5}$ denotes the coefficient of the dimension five operator and $k$ is in the range $0.17\le k\le6.7$. With this in mind, one can obtain LQ masses within the reach of particle colliders with couplings of order $10^{-5}$ to $10^{-3}$. Of course allowing for the above diquark operator will make $\chi$ not a {\it genuine} LQ in the sense that not only the operators in eq.~(\ref{eq2.3}) are present. However, the diquark operators can be suppressed or neglected by imposing a symmetry, in particular a GUT symmetry in a supersymmetric framework. This case has been discussed in~\cite{Dorsner:2009cu} where one embeds $\chi$ in a ${\bf 45}_{H}$-dimensional Higgs representation and has different contractions leading to the quark-lepton interaction and the diquark interaction. Allowing only for the lepton-quark contraction will also lead to the absence of any mixing induced proton decay. In what follows, we will assume that $\chi$ is a {\it genuine} LQ in the sense that either the diquark operator is suppressed or it is altogether absent. In this case, one can assign a lepton and baryon number to $\chi$ such that the accidental lepton and baryon number symmetries of the SM are conserved.

Another aspect relevant in the study of the impact of our model in the UHE neutrino spectrum observed in the IceCube detector is the flavor structure of the interactions in eq.~(\ref{eq2.3}), and its consistency with measurements looking for deviations from the minimal flavor structure of the SM, specially the $2.6\sigma$ deviation from lepton universality presented by the LHCb collaboration on the measurement of $R_{K}$~\cite{Aaij:2014ora}, the ratio between the branching fractions of $B \to K\mu\mu$ and $B \to Kee$.
 In addition, there are hints of LFV reported by the CMS collaboration on the decay $h\to \mu e$~\cite{Khachatryan:2015kon}. The authors in~\cite{Varzielas:2015iva} have used the weak triplet introduced in this work to explain these two measurements by adapting frameworks with non-abelian flavor symmetries that predict the leptonic mixing matrices. Even though the simplest scenarios are those for which the LQ couples to a single generation of leptons, the authors use a data-driven approach to constrain the LQ Yukawa couplings in a generalized scenario using a hierarchical pattern consistent with the observed quark mass pattern. They then analyze various flavor models that lead to different textures of the LQ Yukawa matrix consistent with LFV decays, rare meson decays and lepton universality.

In what follows, we will assume that the LQ couples primarily the first family of quarks to the electron and the muon and the correponding neutrinos. We will also consider the couplings of the LQ to the second and third families of quarks to be suppressed in order to make the collider phenomenology more tractable and, at the same time, to simplify the computation of the rate of UHE neutrino events arising from the LQ component. The study of the viable parameter space consistent with collider constraints and low energy measurements such as LFV decays and atomic parity violation is postponed to secs.~\ref{sec4} and \ref{sec5}.
%%%%%%%%%%%%%%%%%%%%%%%%%%%%%%%%%%%%%%%%%%%%%%%%%%%%%%%%%%%%%%%%%%%
\section{IceCube and PeV Neutrinos}
\label{sec3}
In this section, we study the impact of the model of LQs proposed in sec.~\ref{sec2} in the spectrum of $\mathrm{PeV}$ neutrinos measured at IceCube. In the first place, we revisit the computation of the neutrino-nucleon scattering cross section within the SM, and then we derive the corresponding LQ contribution. The addition of this new physics component leads to particular features in the spectrum, which is studied by computing the expected rate of events. Finally, by adding the rate of events expected from the LQ contribution on top of the SM we determine the masses and couplings that best accommodate the observed spectrum.  
%%%%%%%%%%%%%%%%%%%%%%%%%%%%%%%%%%    
\subsection{Neutrino$\!-\!$nucleon scattering cross section}
\label{sec3.1}
At the IceCube detector, the ultra-high energy (UHE) neutrinos coming from outside the atmosphere are detected by observing the Cherenkov light emitted by the secondary particles produced in their interactions with the nucleons present in the ice. In the standard model (SM), there are charged current (CC) as well as neutral current (NC) neutrino-nucleon interactions, which are mediated by a $W$ or a $Z$ boson, respectively. The topology of the events observed at IceCube depends on the interaction channel as well as on the flavor of the incoming neutrino. The track-like events are induced by CC $\nu_{\mu}$ interactions, while the shower-like events are induced by CC $\nu_{e}$ and $\nu_{\tau}$ interactions and NC interactions of neutrinos of all flavors. \par 
The SM differential cross section for the generic CC interaction $\nu_{\ell}N\to \ell X$, with $\ell = e,\mu,\tau$, $N$ the target nucleon and $X$ the hadronic final state, can be written as,
\beq
\label{eq3.1}
\frac{d^2\sigma}{dx\,dy}^{\!\!\!(CC)}\!\!\!=\,\frac{G^2_F}{\pi}\frac{2M^4_W}{(Q^2+M^2_W)^2}M_NE_{\nu} \,\{xq(x,Q^2)+ x\bar{q}(x,Q^2)(1-y)^2\},
\eeq
where $M_W$ and $M_N$ are the masses of the $W$ and the nucleon respectively, $-Q^2$ is the invariant momentum transferred by the intermediate boson to the hadronic system, and $G_F$ is the Fermi constant. The Bjorken scaling variable $x$ and the inelasticity $y$ used in eq.~(\ref{eq3.1}) are defined as
\beq
\label{eq3.2}
x= \frac{Q^2}{2M_NE_{\nu}y}\qquad\mbox{ and }\qquad y=\frac{E_{\nu}-E_{\ell}}{E_{\nu}},
\eeq
where $E_{\nu}$ and $E_{\ell}$ are the energies carried by the incoming neutrino and by the outgoing lepton in the laboratory frame respectively. Finally, in the case of an isoscalar nucleon $N\equiv (n+p)/2$\footnote{For the numerical computations, we average the nucleon's parton probability distributions using a $5:4$ proton to neutron ratio in ice.}, the quark distribution functions in the differential cross section are given by \cite{Gandhi},
\bea
\label{eq3.3}
q(x,Q^2) &=& \frac{u_v(x,Q^2)+d_v(x,Q^2)}{2}+\frac{u_s(x,Q^2)+d_s(x,Q^2)}{2}+s_s(x,Q^2)+b_s(x,Q^2),\\
\label{eq3.4}
\bar{q}(x,Q^2) &=& \frac{u_s(x,Q^2)+d_s(x,Q^2)}{2}+c_s(x,Q^2)+t_s(x,Q^2),
\eea
where $u,d,s,c,t,b$ denote the distributions corresponding to the various quark flavors in a proton, and the subscripts $v$ and $s$ indicate the valence and sea contributions.\par
Similarly to the CC case, we can write the differential cross section corresponding to the NC process $\nu_{\ell}+N\to \nu_{\ell}+X$ in terms of the variables $x$ and $y$,
\beq
\label{eq3.5}
\frac{d^2\sigma}{dxdy}^{\!\!\!(NC)}\!\!\!=\frac{G^2_F}{2\pi}\frac{M^4_Z}{(Q^2+M^2_Z)^2}M_NE_{\nu}\{ xq^0(x,Q^2)+x\bar{q}^0(x,Q^2)(1-y)^2\},
\vspace*{2mm}
\eeq 
with the following quark distribution functions,
\vspace*{4mm}
\bea
\nonumber
q^0(x,Q^2)&=&\left[\displaystyle \frac{u_v(x,Q^2)+d_v(x,Q^2)}{2}+ \displaystyle \frac{u_s(x,Q^2)+d_s(x,Q^2)}{2}\right](L^2_u+L^2_d)\\[2mm]
\nonumber
&+&\left[\displaystyle \frac{u_s(x,Q^2)+d_s(x,Q^2)}{2}\right](R^2_u+R^2_d)+[s_s(x,Q^2)+b_s(x,Q^2)](L^2_d+R^2_d)\\[2mm]
\label{eq3.6}
&+&[c_s(x,Q^2)+t_s(x,Q^2)](L^2_u+R^2_u),
\eea
\vspace*{2mm}
\bea
\nonumber
\bar{q}^0(x,Q^2)&=&\left[\displaystyle \frac{u_v(x,Q^2)+d_v(x,Q^2)}{2}+ \displaystyle \frac{u_s(x,Q^2)+d_s(x,Q^2)}{2}\right](R^2_u+R^2_d)\\[2mm]
\nonumber
&+&\left[\displaystyle \frac{u_s(x,Q^2)+d_s(x,Q^2)}{2}\right](L^2_u+L^2_d)+[s_s(x,Q^2)+b_s(x,Q^2)](L^2_d+R^2_d)\\[2mm]
\label{eq3.7}
&+&[c_s(x,Q^2)+t_s(x,Q^2)](L^2_u+R^2_u),
\eea
with the quiral couplings given by $L_u=1-(4/3)x_W$, $L_d=-1+(2/3)x_W$, $R_u=-(4/3)x_W$ and $R_d=(2/3)x_W$, where $x_W=\sin^2\theta_W$ is the weak mixing parameter. The CC $\bar{\nu}N$ differential cross section is obtained from eq.~(\ref{eq3.1}) with the contribution $(u_v(x,Q^2)+d_v(x,Q^2))/2$ appearing now in $\bar{q}(x,Q^2)$ instead of $q(x,Q^2)$. Likewise, for the NC $\bar{\nu}N$ differential cross section, the corresponding expression is obtained from eq.~(\ref{eq3.5}) with the replacement $q^0 \leftrightarrow \bar{q}^0$. \par
In addition to their interactions with nucleons, the UHE neutrinos can also interact with electrons in the detection volume. These interactions are proportional to the electron mass and then can be generally neglected compared to the neutrino-nucleon interactions. The only exception is the resonant production of $W^-$ in $\bar{\nu}_e e$ interactions, which occurs at $6.3\,\mathrm{PeV}$. Since this energy is high compared to the most energetic showers observed at IceCube, we will not enter in details regarding the neutrino-electron interactions. The expressions for the differential cross section for these interactions can be found for example in \cite{Gandhi}. \par
\begin{center}
\begin{figure}[H]
\centering
\includegraphics[scale=0.45]{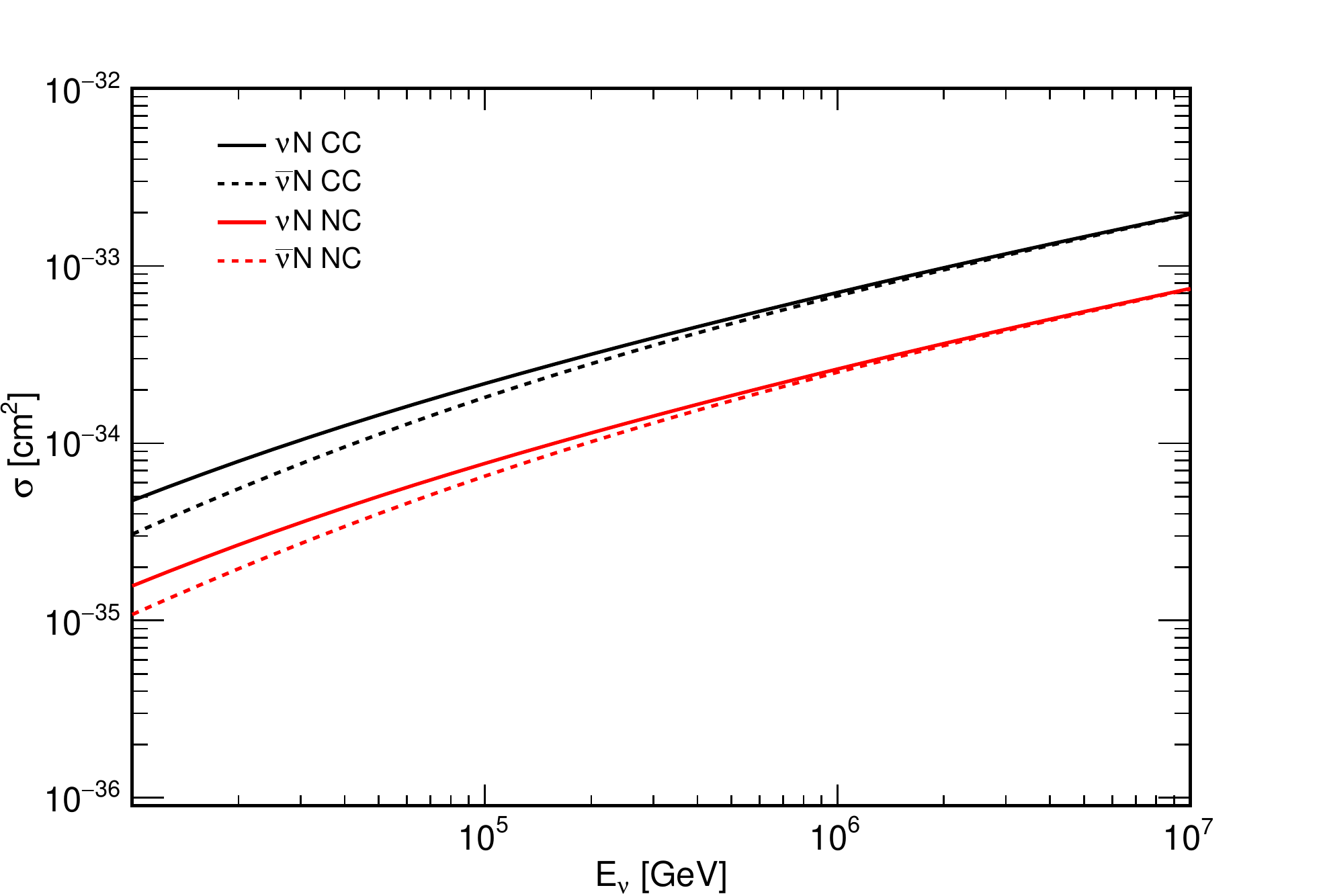}
\caption{Total $\nu N$ and $\bar{\nu} N$ cross sections for the SM CC and NC processes computed using NNPDF2.3 at NLO.}
\label{fig3.1}
\end{figure}
\end{center}

For the numerical computations performed in this paper, we have used the NNPDF2.3 PDF sets \cite{PDFs}. In particular, we use the central values of the PDF sets with $\alpha_s(M_Z)=0.118$ at NLO. The NNPDF2.3 sets provide  a grid division that can go up to $Q^2_{\mathrm{max}}=10^8\,\mathrm{GeV}^2$ in the $Q^2$ axis, and down to $x_{\mathrm{min}}=10^{-9}$ in the $x$ axis. However, given the large uncertainties in the grids for low $x$, we have taken in most of the computations $10^{-6}$ as the lower limit for the $x$-integration. For illustration, we show in figure~\ref{fig3.1} the total $\nu N$ and $\bar{\nu} N$ cross sections in terms of the incoming neutrino energy $E_{\nu}$ for the SM contributions.
\par
\begin{center}
\begin{figure}[H]
\centering
%\hspace*{-0.4cm}
\subfloat{\includegraphics[scale=0.40]{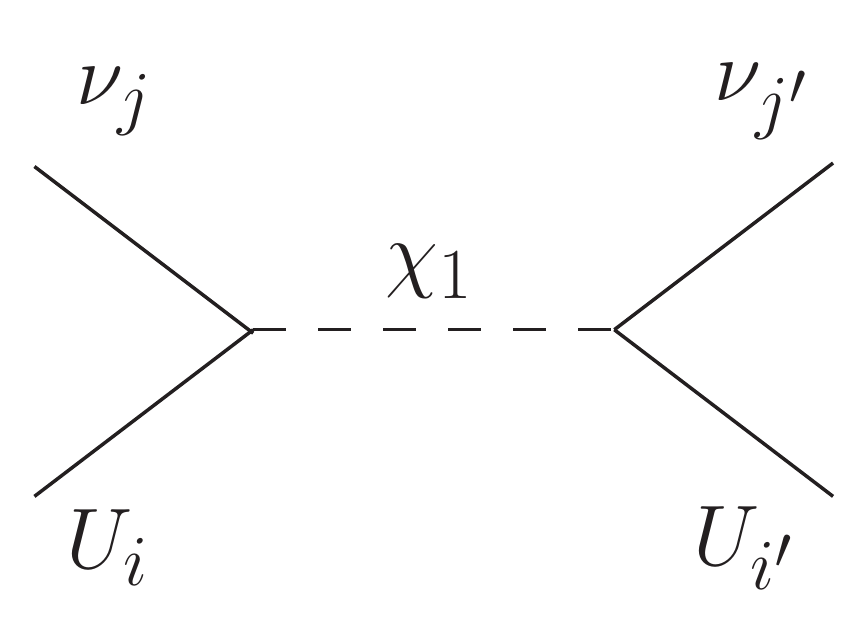}}
\hspace*{0.2\textwidth}
%\label{fig1a}}
\subfloat{\includegraphics[scale=0.40]{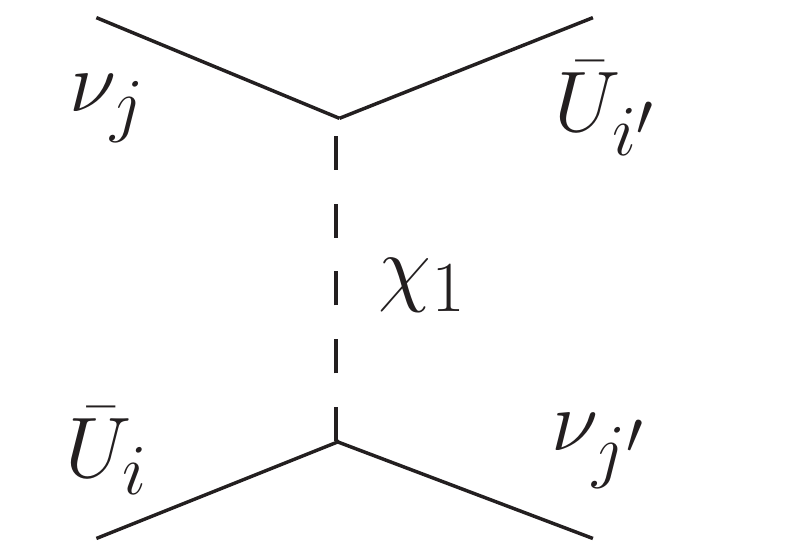}}\\
%\label{fig1b}}
\subfloat{\includegraphics[scale=0.40]{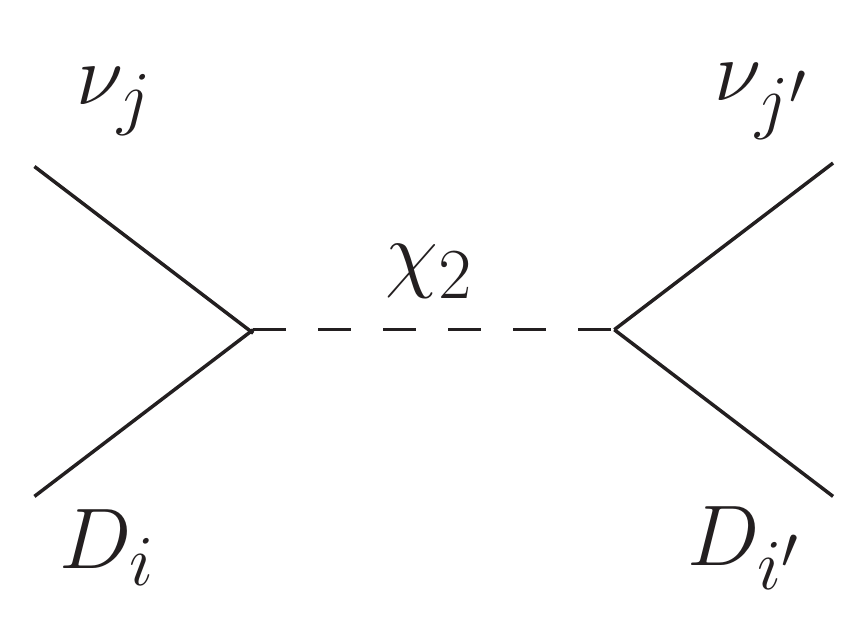}}
\hspace*{0.2\textwidth}
%\label{fig1a}}
\subfloat{\includegraphics[scale=0.40]{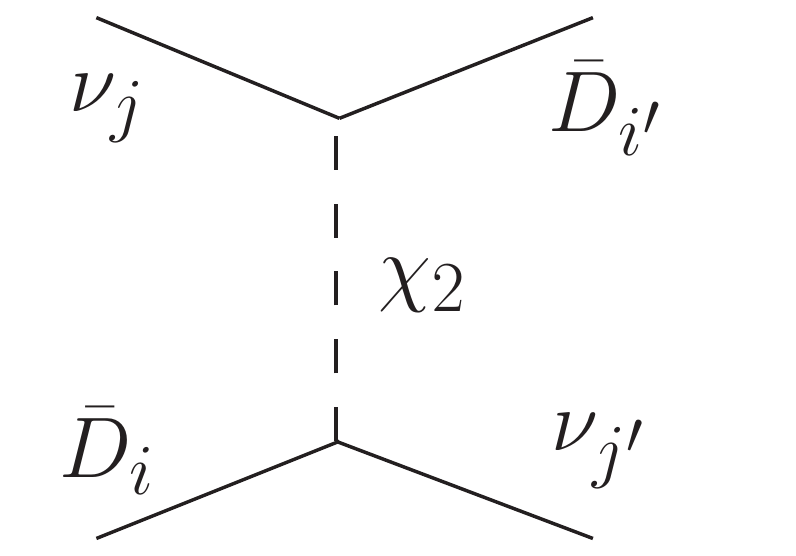}}\\
\subfloat{\includegraphics[scale=0.40]{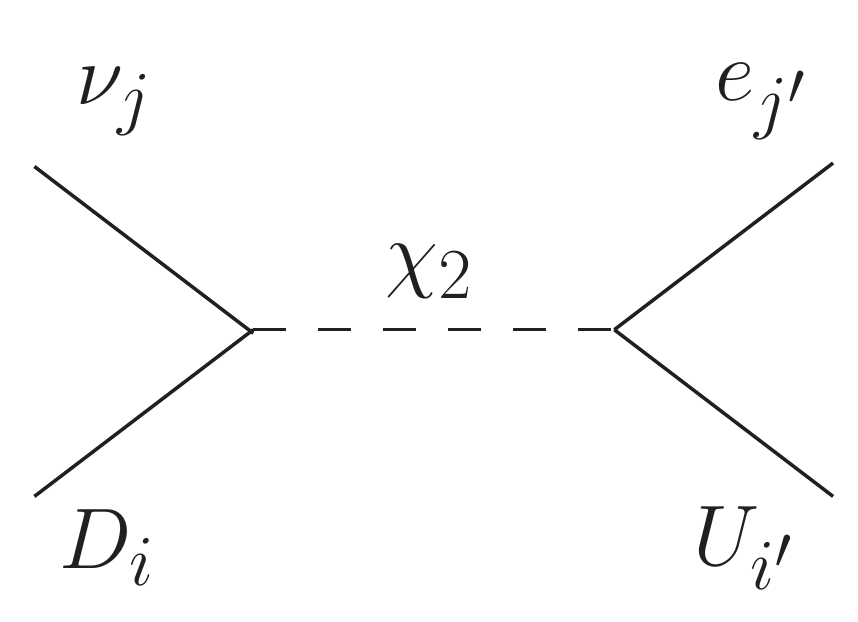}}
\hspace*{0.2\textwidth}
%\label{fig1a}}
\subfloat{\includegraphics[scale=0.40]{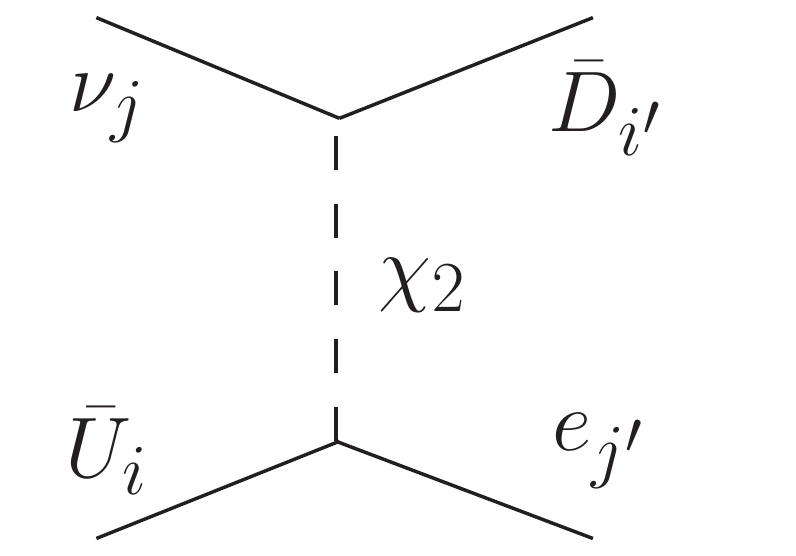}}
\caption{Feynman diagrams contributing to the $\nu N$ interaction. The two first rows correspond to NC processes and the last one to CC processes.}
\label{fig3.2}
\end{figure}
\end{center}

In order to study the impact of the proposed LQ in the energy distribution of the events expected at IceCube, we must compute the LQ contribution to the neutrino-nucleon cross section. From eq.~(\ref{eq2.3}), we see that only $\chi_1$ and $\chi_2$ give contributions to this cross section; the analogue to the SM's NC processes are provided by both $\chi_1$ and $\chi_2$, whereas the final states corresponding to the SM's CC processes are produced only through $\chi_2$. The corresponding Feynman diagrams are depicted in figure~\ref{fig3.2}, where $U$ and $D$ denote up- and down-type quarks, and the indices $i,i^{\prime}$ and $j,j^{\prime}$ indicate the number of family for quarks and leptons respectively.
%%%%
%%%%
\par
\begin{center}
\begin{figure}[H]
\centering
\hspace*{1mm}
\subfloat{\includegraphics[scale=0.42]{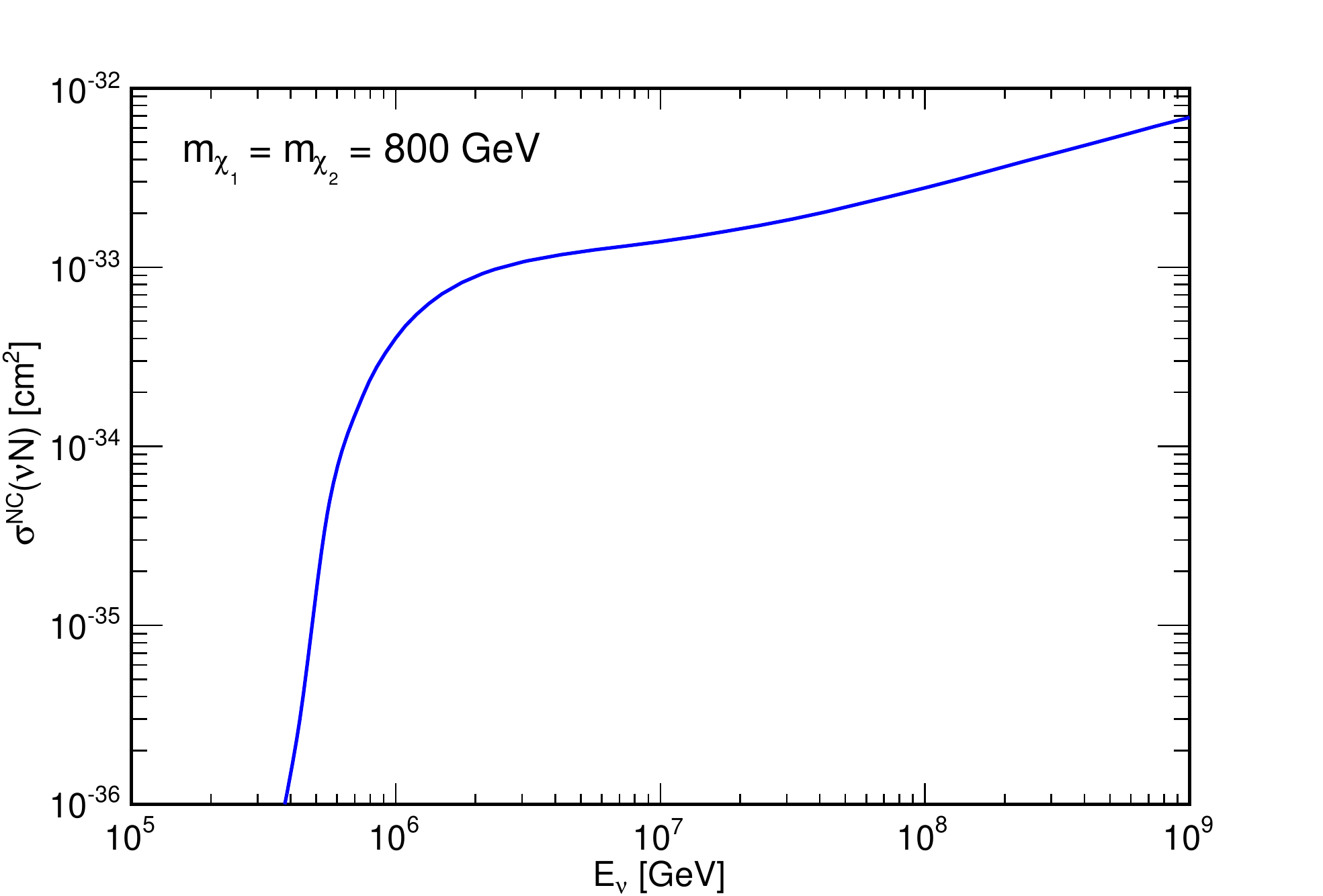}}
\hspace*{1mm}
\label{fig3.3a}
\subfloat{\includegraphics[scale=0.42]{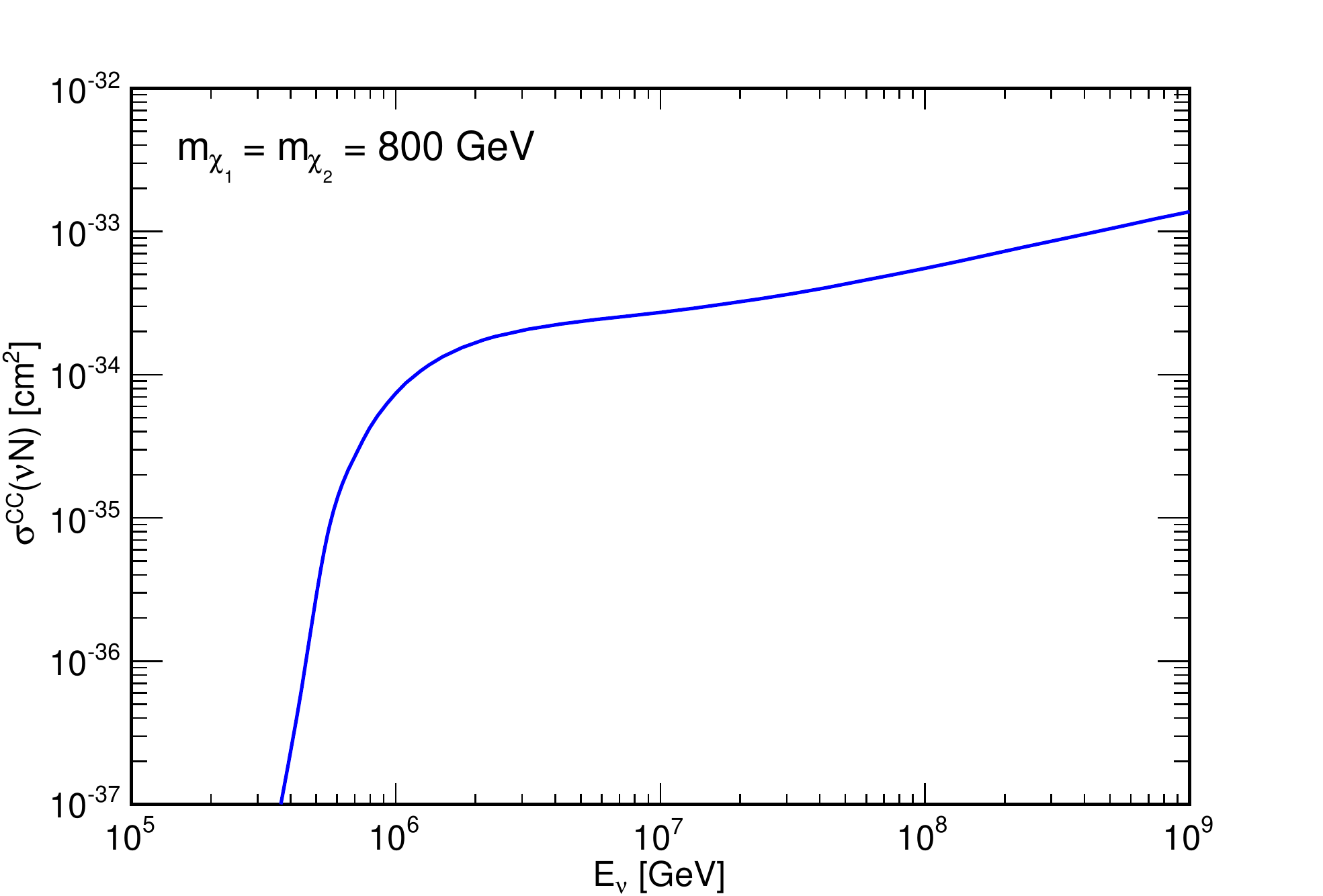}}
\label{fig3.3b}
\caption{Neutrino-nucleon scattering cross section for NC (left) and CC (right) processes as a function of the incoming neutrino energy $E_{\nu}$. In both cases, the cross sections corresponding to $\nu_{e}$ and $\nu_{\mu}$ have been added.}
\label{fig3.3}
\end{figure}
\end{center}

Since the cross sections corresponding to the $s$-channel processes shown in the first column of figure~\ref{fig3.2} are resonance enhanced, we will assume that they dominate the neutrino-nucleon interaction and use the narrow width approximation for both $\chi_1$ and $\chi_2\,$\footnote{In addition to the contribution of the s-channel diagrams to the NP amplitude, there are also contributions arising from the u-channel diagrams depicted in figure~\ref{fig3.2} as well as from the interference between the LQ and the SM amplitudes. To compare these different contributions, we have computed $\sigma (\nu N)$ as a function of $E_{\nu}$ inclusively for various values of the couplings $\lambda$. In all the cases, the cross section computed using only the s-channel diagrams exhibits the same behaviour than that computed using all the contributions, with the differences being of the order of the PDFs uncertainties. Taking this analysis into account we neglect both the  contributions from the u-channel diagrams and any interference effect with the SM amplitude. Additionally, this approach greatly simplifies the statistical analysis of the IceCube data.}. Also, we consider a scenario in which the LQ triplet couples only to the first generation of quarks and, for the sake of simplicity, to the first and second generations of leptons; hence, we have $\lambda^i_{j}=0$ for $i\neq 1$ and/or $j =3$. Within this scenario, the differential cross section for the NC and CC processes can be written as follows
\bea
\label{eq3.8}
\frac{d\sigma}{dy}(\nu_{j} N \to \nu_{j^{\prime}}\,X)&=&\frac{|\lambda^1_{j}|^2|\lambda^1_{j^{\prime}}|^2}{32s}\left\{\frac{m_{\chi_1}}{\Gamma_{\chi_1}}f_{u}(m^2_{\chi_1}/s,m^2_{\chi_1}y)+\frac{m_{\chi_2}}{4\Gamma_{\chi_2}}f_{d}(m^2_{\chi_2}/s,m^2_{\chi_2}y) \right\},\\
\label{eq3.9}
\frac{d\sigma}{dy}(\,\nu_{j} N \to e^-_{j^{\prime}}\,X)&=&\frac{|\lambda^1_{j}|^2|\lambda^1_{j^{\prime}}|^2}{128s}\frac{m_{\chi_2}}{\Gamma_{\chi_2}} f_{d}(m^2_{\chi_2}/s,m^2_{\chi_2}y)\,,
\eea
where $j,j^{\prime}=1,2$, $\Gamma_{\chi_{1}}$ and $\Gamma_{\chi_{2}}$ are the total widths of $\chi_{1,2}$, $s=2 M_N E_{\nu}$ is the center-of-mass energy squared, and $f_{u,d}$ are the distribution functions of the up and down quarks in the nucleon, respectively. In the case of an isoscalar nucleon, these functions turns out to be equal and given by,
\beq
\label{eq3.10}
f_u(x,Q^2)=f_d(x,Q^2)=\frac{u_v(x,Q^2)+d_v(x,Q^2)}{2}+\frac{u_s(x,Q^2)+d_s(x,Q^2)}{2}.
\eeq  
In eqs.~(\ref{eq3.8}) and (\ref{eq3.9}), the fractional momentum $x$ has been integrated out by using the narrow width approximation for both $\chi_1$ and $\chi_2$. As a consequence of this, the distribution functions are evaluated at $x=m^2_{\chi_{1,2}}/s$ and $Q^2=xys=m^2_{\chi_{1,2}}y$. In order to compute the widths of $\chi_1$ and $\chi_2$ we assume that $\Gamma_{\chi_1}$ is saturated by the decay $\chi_1\to \nu_{e,\mu}\,u$ while $\Gamma_{\chi_2}$ is saturated by $\chi_2\to \nu_{e,\mu}\,d$ and $\chi_2\to \ell(=e,\mu)\,u$. Thus, these widths are written in terms of the couplings as follows
\bea
\label{eq3.11}
\Gamma_{\chi_1}&=&\frac{m_{\chi_1}}{16\pi}(|\lambda^1_1|^2+|\lambda^1_2|^2), \\
\label{eq3.12}
\Gamma_{\chi_2}&=&\frac{m_{\chi_2}}{16\pi}(|\lambda^1_1|^2+|\lambda^1_2|^2). 
\eea 
By combining eqs.~(\ref{eq3.11}) and (\ref{eq3.12}) with eqs.~(\ref{eq3.8}) and
 (\ref{eq3.9}), the differential cross sections for NC and CC $\nu N$ scattering are expressed in terms of the couplings $\lambda^1_1$ and $\lambda^1_2$ and the LQ masses
  $m_{\chi_1}$ and $m_{\chi_2}$. In the case of antineutrino-nucleon scattering the expressions for the NC and CC processes are the same but with the quark distributions replaced by the respective antiquark distributions, i.e.,  $f_u\to f_{\bar{u}}$ and $f_d\to f_{\bar{d}}$. In figure~\ref{fig3.3} we show the LQ contribution to the total $\nu N$ cross section for the NC and CC processes. For concreteness, we have considered the case in which $\chi_1$ and $\chi_2$ are degenerate in mass with $m_{\chi_1}=m_{\chi_2}=800\,\mathrm{GeV}$ and the LQ couplings are such that $|\lambda^1_1|=|\lambda^1_2|=1$. From figure~\ref{fig3.3} and comparing with the SM cross sections in figure~\ref{fig3.1}, we see that the LQ contribution turns on when the incoming neutrino energy is enough to produce the resonances $\chi_{1,2}$. This occurs when the center-of-mass energy is such that $\sqrt{s}\geqslant m_{\chi_{1,2}}$ or, equivalently, when $E_{\nu}\geqslant m^2_{\chi_{1,2}}/2M_N $.
%\vspace*{-10mm}  
%\vspace*{-1cm}
In contrast to the SM cross section, the total cross section induced by the resonant LQ production is higher for the NC reactions, which involve both $\chi_1$ and $\chi_2$, than for the CC reactions, which proceed only via $\chi_2$. In the case of non-degenerate masses, we note that the splitting in mass cannot be greater than $\sim\!50\,\mathrm{GeV}$ due to the constraints arising from the oblique parameters~\cite{Dorsner:2016wpm,EWPD}. For such small difference in mass between $\chi_1$ and $\chi_2$, the behaviour of the cross section with respect to the incoming neutrino energy is quite similar, with the actual values being slightly smaller or higher depending on wether the mass of $\chi_1$ or $\chi_2$ is increased or decreased with respect to the degenerate case. In what follows,  we will focus then in the case $m_{\chi_1}=m_{\chi_2}$.
%\vspace*{-4mm}
%%%%%%%%%%%%%%%%%%%%%%%%%%%%%%%%%%%%%%%%%%%%%
\subsection{Event rate at IceCube and LQ contribution}
\label{sec3.2}
The LQ contribution to the total number of events induced by neutrinos of a certain flavor can be written as follows,
\beq
\label{eq3.13}
N  = T\cdot \Omega\cdot \int_0^{\infty}dE_{\nu}\,N_{\mathrm{eff}}\frac{d\phi}{dE_{\nu}}\int_0^1 dy\frac{d\sigma}{dy},\\[2mm]
\eeq
where $T$ is the exposure time, $\Omega$ the solid angle of coverage, $N_{\mathrm{eff}}$ the effective number of target nucleons, $d\phi/dE_{\nu}$ is the flux of the incoming neutrinos and $d\sigma/dy$ is the differential neutrino-nucleon cross section corresponding to the sum of eqs.(\ref{eq3.8}) and (\ref{eq3.9}). From eq.~(\ref{eq3.13}), the distribution of the number of events with respect to the incoming neutrino energy and the inelasticity is
\beq
\label{eq3.14}
\frac{dN}{dE_{\nu}dy}=T\, \Omega\, N_{\mathrm{eff}}\frac{d\phi}{dE_{\nu}}\frac{d\sigma}{dy}.
\eeq
We note that, in order to compare with the rate of events observed at IceCube, we must use the distribution of the number of events with respect to the deposited energy $E$, which is always smaller than the incoming neutrino energy $E_{\nu}$. The predicted number of events due to the LQ contribution in the deposited energy interval $\Delta \equiv(E_i,E_f)$ is given by
\beq
\label{eq3.15}
N_{\Delta}=\int^1_0 \int^{E_f}_{E_i} dydE \frac{dN}{dEdy}=\int^1_0 \int^{E_f}_{E_i}dydE\frac{dN}{dE_{\nu}dy}\frac{dE_{\nu}}{dE}=T\cdot \Omega \cdot \int^1_0 \int^{E_f}_{E_i} dydE \,N_{\mathrm{eff}}\frac{d\sigma}{dy}\frac{d\phi}{dE_{\nu}}\frac{dE_{\nu}}{dE}. 
\eeq  
By changing variables from the deposited energy $E$ to the incoming neutrino energy $E_{\nu}$ in eq.~(\ref{eq3.15}), we obtain
\beq
\label{eq3.16}
N_{\Delta}= T\cdot \Omega \cdot \int^1_0 \int^{E^f_{\nu}(E_f,y)}_{E^i_{\nu}(E_i,y)} dydE_{\nu} \,N_{\mathrm{eff}}\frac{d\sigma}{dy}\frac{d\phi}{dE_{\nu}} . 
\eeq
The relation between the deposited energy and the incoming neutrino energy depends on the interaction channel. In this study we follow the approach used in ref.~\cite{Soni}, which we summarize in the following. For NC events, the outgoing hadrons carry an energy $E_X=yE_{\nu}$, and the corresponding deposited energy is given by $E_{\mathrm{had}}=F_XyE_{\nu}$, where $F_X$ is the ratio of the number of photo-electrons yielded by the hadronic shower to that produced by an equivalent-energy electromagnetic shower. This quantity is parameterized as \cite{shower2,Soni}
\beq
\label{eq3.17}
F_X=1-\left(\frac{E_X}{E_0}\right)^{-m}(1-f_0),
\eeq
where $E_0=0.399~\mathrm{GeV}$, $m=0.130$ and $f_0=0.467$ are the best-fit values obtained from simulations in ref.~\cite{shower2}. The energy carried by the final state neutrino is missed and thus the total deposited energy for NC $\nu_e$- and $\nu_{\mu}$-events is $E_{\mathrm{NC}}=F_XyE_{\nu}$. In the case of CC events, in contrast, the energy of the final state lepton, $E_{e,\mu}=(1-y)E_{\nu}$, is completely deposited giving rise to a total deposited energy given by $E_{\mathrm{CC}}=E_{e,\mu}+E_{\mathrm{had}}$. The remaining ingredients appearing in eq.~(\ref{eq3.16}) are set in the following manner:
\begin{itemize}
\item For the time of exposure we take $T=1347$ days, corresponding to four years of IceCube data between $2010$ and $2014$ \cite{Aartsen:2015zva}.
\item The solid angle of coverage is $\Omega = 2\pi\,\mathrm{sr}$ for events coming from the southern hemisphere (downward-going neutrino events). Due to attenuation effects in the Earth, the effective solid angle for northern events turns out to be smaller by a shadow factor that depends on $E_{\nu}$. This factor can be written as \cite{Gandhi}
\beq
\label{eq3.18}
S(E_{\nu})=\frac{1}{2\pi}\int^{0}_{-1}d\cos\theta\int d\phi \exp[-z(\theta)/L_{\mathrm{int}}(E_{\nu})],
\eeq
where the function $z(\theta)$ gives the thickness of the Earth as a function of the angle of incidence of the incoming neutrinos and $L_{\mathrm{int}}(E_{\nu})$ is the interaction length, which depends on the flavor of the incoming neutrino. Thus, for an isotropic neutrino flux, the total solid angle of coverage is given by $\Omega_{\mathrm{tot}}= 2\pi (1+S(E_{\nu}))\,\mathrm{sr}$, from which we see that for a fully opaque Earth $\Omega_{\mathrm{tot}}= 2\pi\,\mathrm{sr}$, while for a transparent Earth, $\Omega_{\mathrm{tot}}= 4\pi\,\mathrm{sr}$. The LQ contribution modify in principle the interaction length and therefore the shadow factor. However, the deviation from the SM expectation for the total solid angle turns out to be small, so that the LQ contribution can be neglected in the computation of the shadow factor. On the other hand, $S(E_{\nu})$ is a monotonically decreasing function of the incoming energy $E_{\nu}$ that, in the range of energies relevant at Icecube ($10\,\mathrm{TeV}-10^4\,\mathrm{TeV}$), varies between $1$ and $\sim\!0.15$. For simplicity, we will cosider the total solid angle of coverage as a constant and present in the following the results for both limiting cases mentioned above (for further details on attenuation effects in the Earth see App.~\ref{A1}).
\item The effective number of target nucleons depends on the energy of the incident neutrinos, $N_{\mathrm{eff}}=N_AV_{\mathrm{eff}}(E_{\nu})$, where $N_A=6.022\times 10^{23}\,\mathrm{cm}^{-3}$ water equivalent (we) is Avogadro's number. The effective target volume can  be written as $V_{\mathrm{eff}}(E_{\nu})=M_{\mathrm{eff}}/\rho_{\mathrm{ice}}$ with $M_{\mathrm{eff}}$ the effective target mass and $\rho_{\mathrm{ice}}$ the density of ice. The effective target mass increases with $E_{\nu}$ and reaches a maximum value of  $\simeq 400\,\mathrm{Mton}$ above $100\,\mathrm{TeV}$, in the case of $\nu_e$ CC events, and above $1\,\mathrm{PeV}$ for NC events and CC events induced by $\nu_{\mu}$ and $\nu_{\tau}$ \cite{icecubeNeff}. For the computation of the LQ contribution to the event rate observed at IceCube, we use the maximum value for $M_{\mathrm{eff}}$ which corresponds to $V_\mathrm{eff}=0.44\,\mathrm{km}^3$we. 
\item For each neutrino flavor $i$, we assume an isotropic, single power-law flux that is parameterized as follows,
\beq
\label{eq3.19}
\frac{d\phi_i}{dE_{\nu}}= \phi_0f_i \left(\frac{E_{\nu}}{100~\mathrm{TeV}}\right)^{-\gamma},
\eeq  
where $f_i$ is the fraction of neutrinos of the $i$-th flavor at Earth, $\gamma$ is the power law spectral index and $\phi_0$ is the all-flavor neutrino flux at $100~\mathrm{TeV}$. We use the most commonly considered scenario in which the flux is dominated by the decay of pions and their daugther muons giving rise to a flavor ratio of $(1/3,2/3,0)$ at source. This ratio tends to equalize at Earth due to neutrino oscillations averaged over astronomical distances. Hence, in eq.~(\ref{eq3.19}), we set $f_i=1/3$ for $i=e,\mu,\tau$. Also, an equal $\nu$ and $\bar{\nu}$ flux is used \cite{icecube1}. Regarding the spectral parameters $\phi_0$ and $\gamma$, we take the best-fit values obtained in ref.~\cite{icecubeLL} by performing a maximum-likelihood combination of the results from six different IceCube searches. The spectral parameters resulting from this analysis in the case of the single-power law are given by
\beq
\label{eq3.20}
\phi_0 = (6.7^{+1.1}_{-1.2})\times 10^{-18}~\mathrm{GeV}^{-1}\mathrm{s}^{-1}\mathrm{sr}^{-1}\mathrm{cm}^{-2},
\eeq 
\vspace*{-8mm}
\beq
\label{eq3.21}
\gamma = 2.50 \pm 0.09 \,.
\eeq
\end{itemize}
\par
\begin{center}
\begin{figure}[H]
\centering
\hspace*{0.4cm}
\subfloat{\includegraphics[scale=0.41]{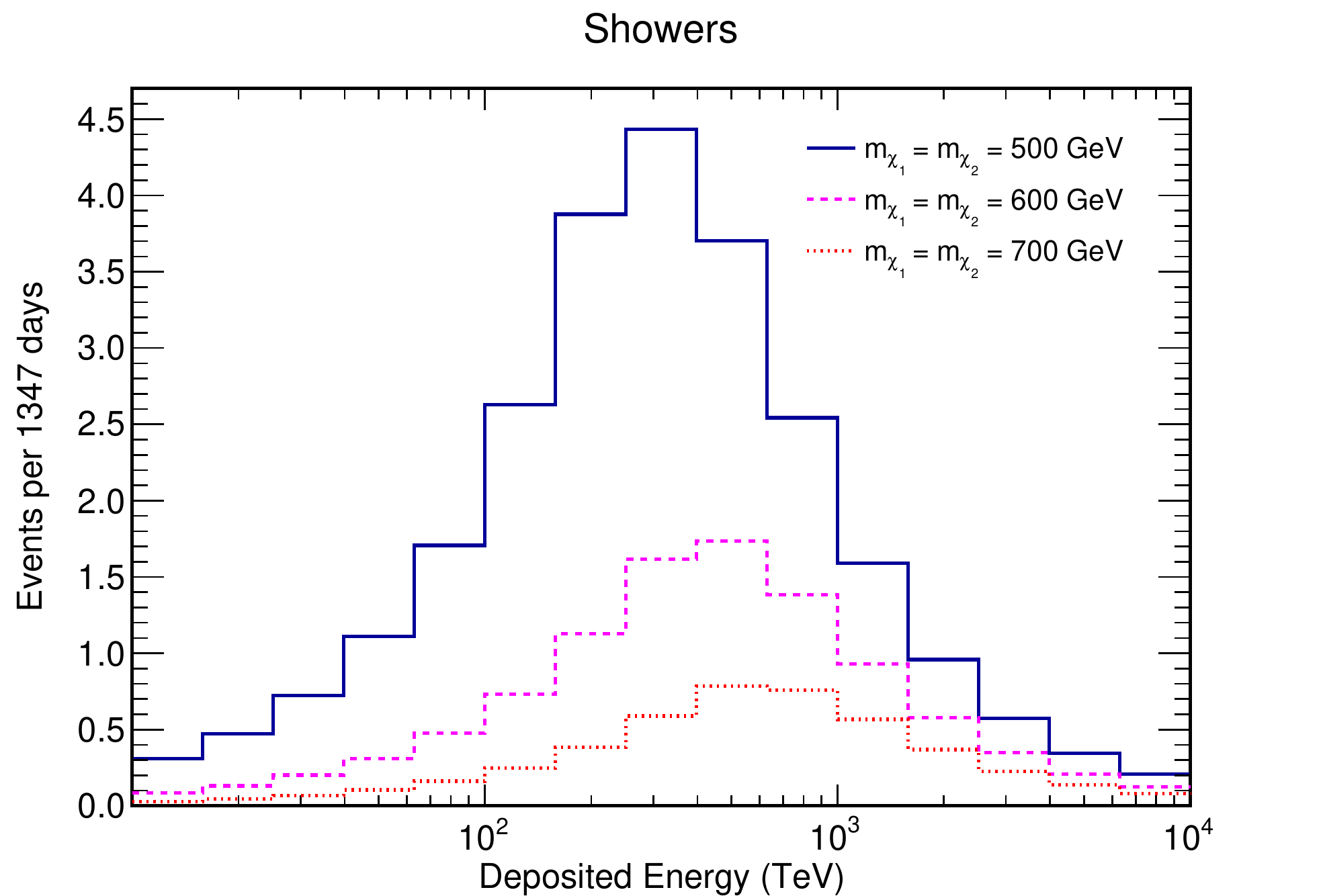}}
%\hspace*{0.2\textwidth}
%\label{fig1a}}
\subfloat{\includegraphics[scale=0.41]{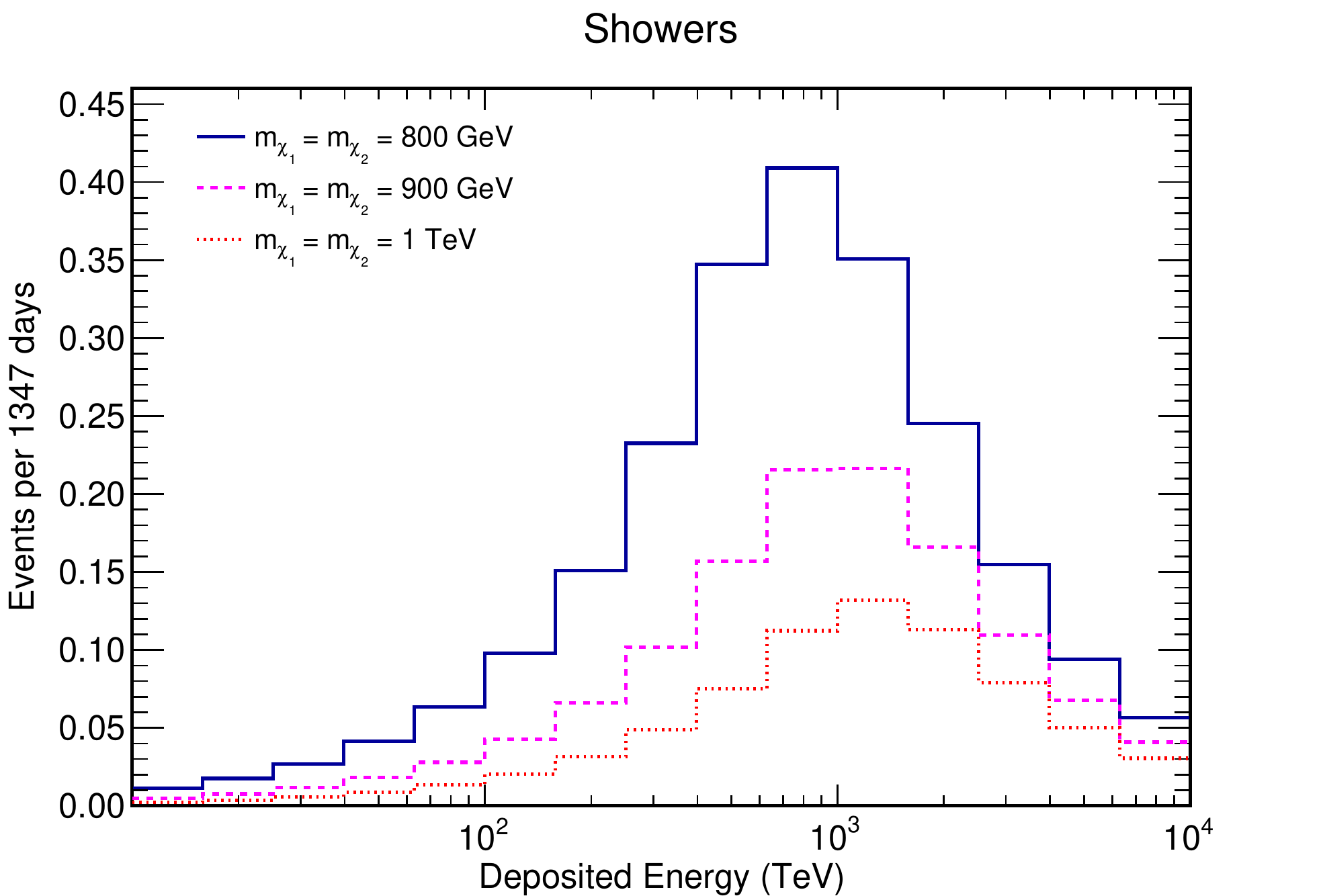}}\\[1mm]
\hspace*{0.4cm}
\subfloat{\includegraphics[scale=0.41]{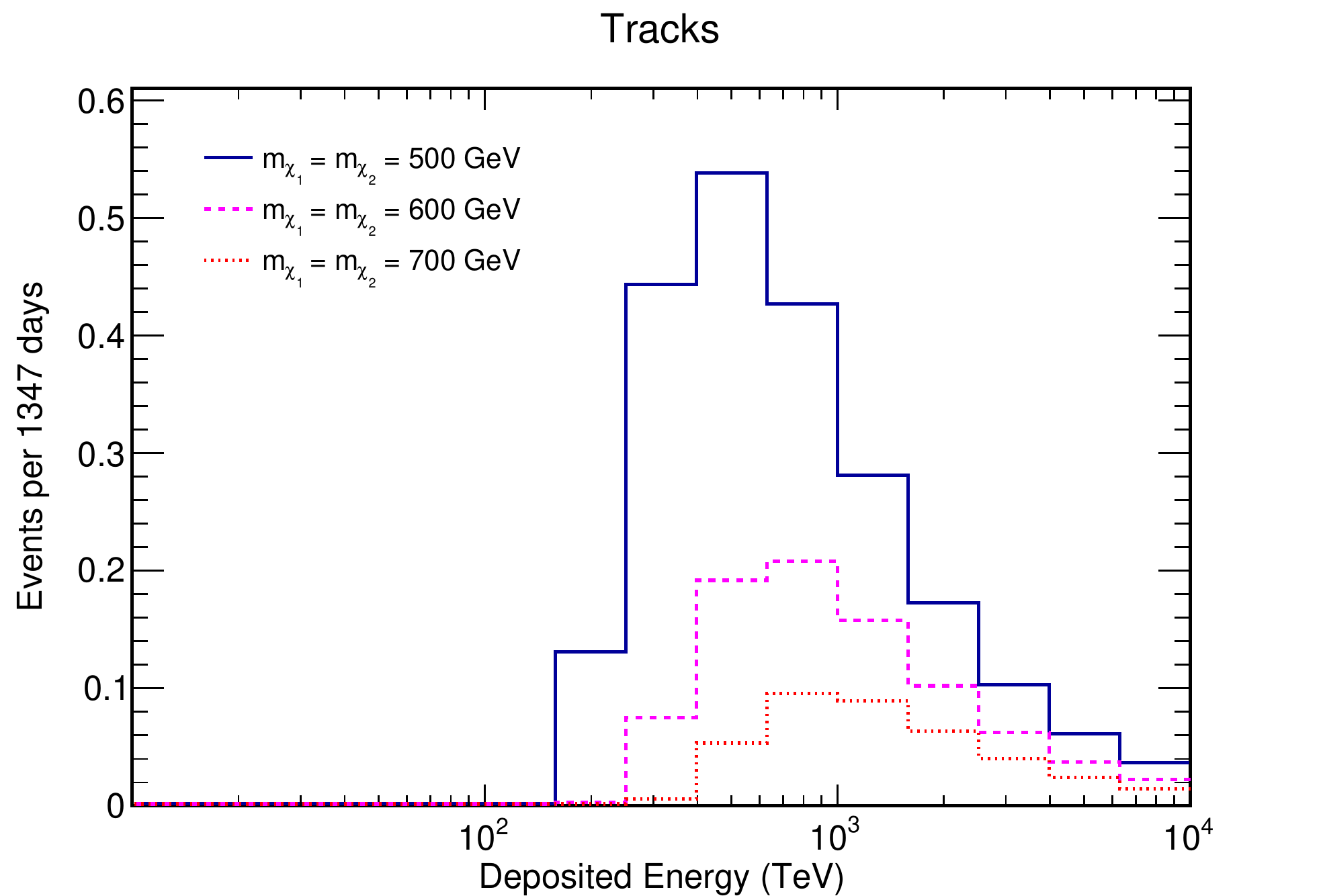}}
\subfloat{\includegraphics[scale=0.41]{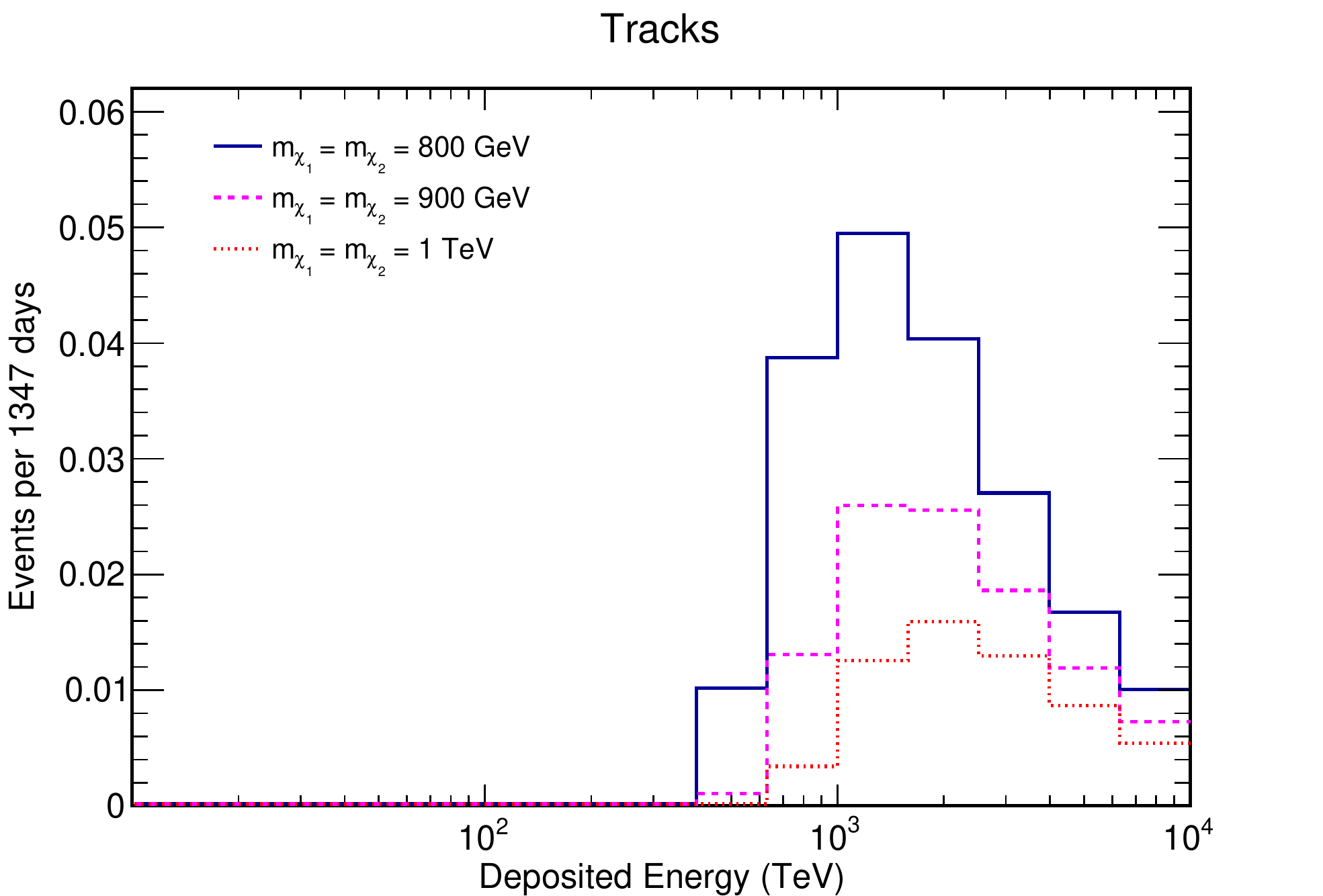}}\\[1mm]
\hspace*{0.4cm}
\subfloat{\includegraphics[scale=0.41]{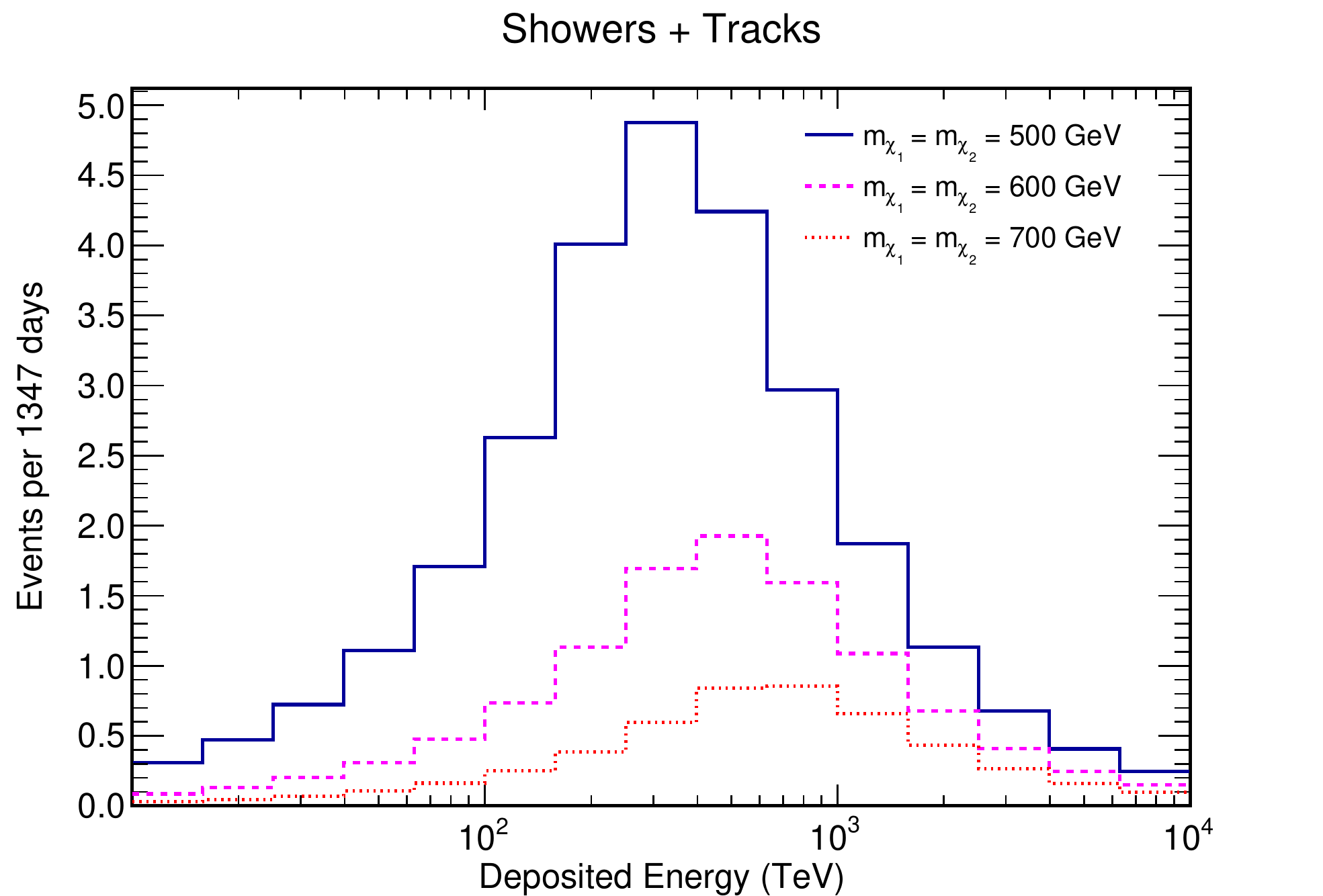}}
\subfloat{\includegraphics[scale=0.41]{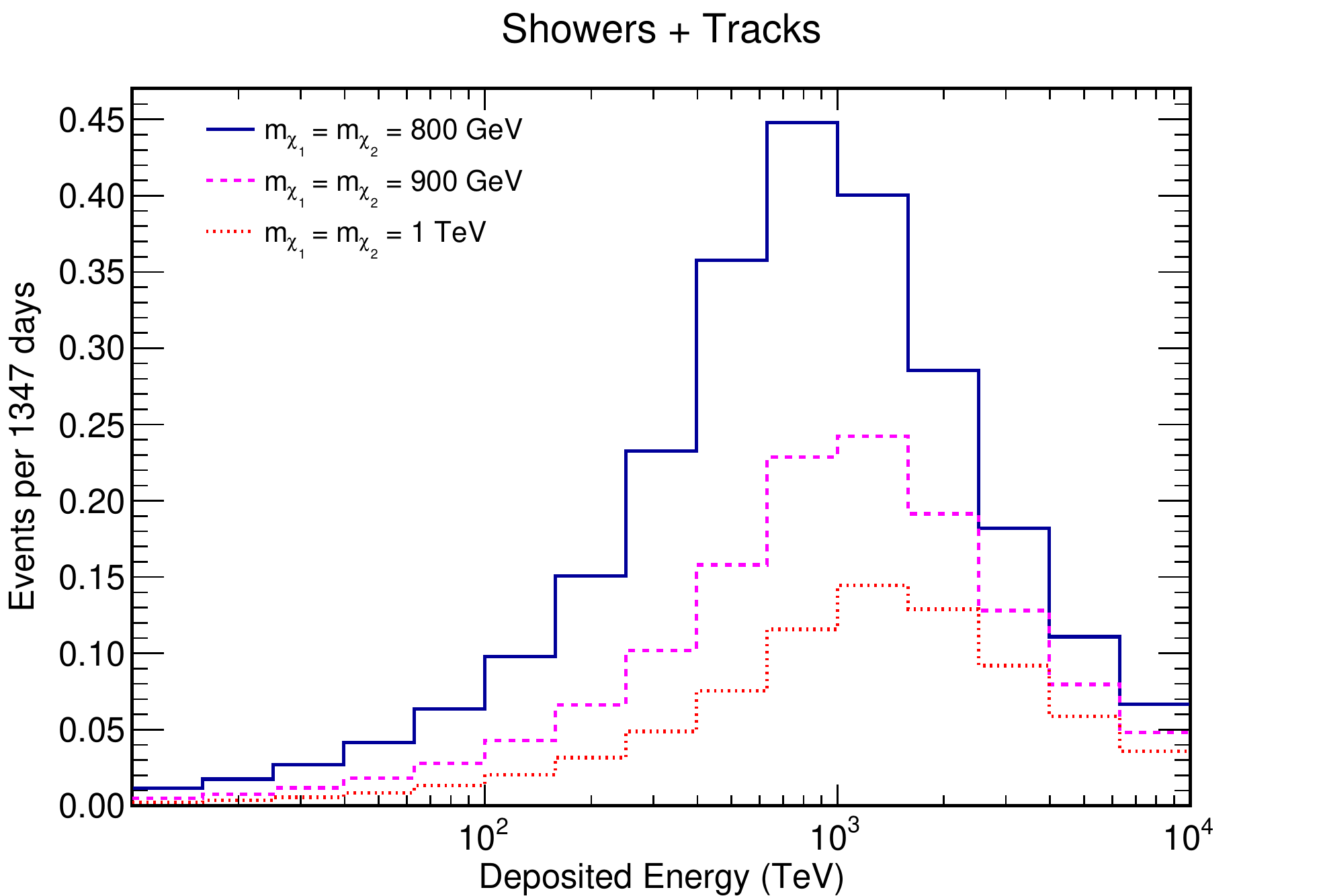}}
\vspace*{1mm}
\caption{Number of events expected from the LQ contribution as a function of the deposited energy for different values of $m_{\chi_1}$ (= $m_{\chi_2}$). The upper and middle panels correspond to shower-and track-like events, respectively, and the lower panels display the total number of events.} 
\label{fig3.4}
\end{figure}
\end{center}

In order to illustrate the LQ component expected for the number of events, we have applied eq.~(\ref{eq3.16}) to $15$ bins of deposited energy in the range $[10\,\mathrm{TeV},10\,\mathrm{PeV}]$. In figure~\ref{fig3.4} we show the LQ component to shower- and track-like events along with their sum for $|\lambda^1_1|=|\lambda^1_2|=1$ and for different values of the LQ masses ranging between $500\,\mathrm{GeV}$ and $1\,\mathrm{TeV}$. We note that, in our scenario, the LQ contributes to shower-like events via NC processes initiated by $\nu_{e,\mu}$ (and $\bar{\nu}_{e,\mu}$) or CC $\nu_e N$ (and $\bar{\nu}_eN$) interactions; in the case of track-like events, the LQ contribution arises only from the CC process $\nu_{\mu}N\to \mu^{-}X$ (and $\bar{\nu}_{\mu}N\to \mu^{+}X$). An important feature of the distributions in figure~\ref{fig3.4} is that the regions of deposited energy at which they peak increase with the LQ mass. This general behaviour is inherited from the distribution of the number of events with respect to $E_{\nu}$. On the other hand, due to the fact that NC and CC processes deposit different ammounts of energy, the distributions of track-like events exhibit the threshold at $m^2_{\chi}/2M_N$ for resonant production of $\chi_2$ (middle panels of figure~\ref{fig3.4}), while those corresponding to shower-like events keep different from zero for all the considered bins (upper panels of figure~\ref{fig3.4}).\footnote{In the case of NC events (induced by $\nu_{e,\mu}$), any value of $E_{\nu}$ can contribute to a certain bin of deposited energy $\Delta=[E_i,E_f]$ providing $E_{\nu}>E_i$. In contrast, for CC events, only values of $E_{\nu}$ within $\Delta$ or slightly above $E_f$ contribute.}
%\vspace*{-6.6mm}
%%%%%%%%%%%%%%%%%
%%%%%%%%%%%%%%%%%
\vspace*{-2mm}
\par
In the following subsection, we add the LQ contribution in top of the spectrum expected from the SM + background hypothesis and study its implications on the spectrum actually observed by IceCube.   
\subsection{Statistical Analysis and Results}
\label{sec3.3}
We consider the number of events in the $i$-th bin of deposited energy, $n_i$, as a poisson variable and parameterize the respective expected number of events as 
\beq
\label{eq3.22}
\nu_i = \mu \,y^{s}_i + y^{b}_i,
\eeq
where $\mu \equiv |\lambda^1_1|^2+|\lambda^1_2|^2$, $\mu\,y^{s}_i$ is the number of events arising from the LQ contribution and $y^{b}_i$ the number of events expected from the SM + background model. The values $y^{s}_i$ were computed by using eq.~(\ref{eq3.16}), whereas the $y^{b}_i$'s were taken from ref.~\cite{Aartsen:2015zva}. In order to estimate the $\mu$ parameter, we minimize the following statistic test,
\beq
\label{eq3.23}
\chi^2(\mu)\equiv -2\,\mathrm{ln}(L(\mu))=-2\sum_i \mathrm{ln} \left(\frac{\nu^{n_i}_i e^{-\nu_i}}{n_i!}\right)=2\sum_i (\nu_i - n_i\mathrm{ln}(\nu_i)+\mathrm{ln}(n_i!)),
\eeq
where $L(\mu)$ is the likelihood function. We note that the minimization of $\chi^2(\mu)$ is equivalent to the maximization of the likelihood function and that the last term of eq.~(\ref{eq3.23}) can be dropped during the minimization. The results obtained for LQ masses between $500\,\mathrm{GeV}$ and $1.5\,\mathrm{TeV}$ are shown in table \ref{table3.1}, where $\hat{\mu}$ denotes the value of $\mu$ that minimizes $\chi^2(\mu)$, and where the cases $\Omega = 4\pi$ and $2\pi~\mathrm{sr}$ have been considered. \par
For the two lowest masses considered here, namely $500$ and $600~\mathrm{GeV}$, we obtain non-physical values for $\hat{\mu}$. This result can be understood from the fact that the event distributions corresponding to these masses peak in the region between $100$ and $1000~\mathrm{TeV}$, where the SM + background prediction is already above the observed spectrum. For LQ masses between $700$-$1200~\mathrm{GeV}$, we obtain increasing positive values of $\hat{\mu}$, which is in fact expected since the LQ contribution deacreases with increasing $m_{\chi}$ (see figure~\ref{fig3.4}). Finally, for $m_{\chi}>1200~\mathrm{GeV}$, the maximum of the corresponding event distributions lies at the right end or even beyond the range of deposited energy considered at IceCube. Accordingly, LQs with these masses contribute mainly to the most energetic bins in which no event have been observed, forcing the $\hat{\mu}$'s to decrease and even to become negative for $m_{\chi}=1500~\mathrm{GeV}$.\par
We have also used the estimates in table \ref{table3.1} to obtain upper limits for the parameter $\mu$. In this case we use the following statistic for testing values of $\mu$ such that $\mu \geq \hat{\mu}$ \cite{Cousins,Cowan},
\beq
\label{eq3.24}
q_{\mu}\equiv -2\mathrm{ln}(\lambda(\mu))=-2\mathrm{ln}\left(\frac{L(\mu)}{L(\hat{\mu})}\right)= \chi^2(\mu)-\chi^2(\hat{\mu})=-2\sum_i (\hat{\nu}_i - \nu_i + n_i(\mathrm{ln(\nu_i)-ln(\hat{\nu}_i)})),
\eeq
where $\lambda(\mu)$ is the profile likelihood ratio, and $\hat{\nu}_i$ is obtained from eq.~(\ref{eq3.22}) with the replacement $\mu\to \hat{\mu}$. Then, the $95\% \,\mathrm{CL}$ upper limit is defined as the maximum value of $\mu$ for which $p_{\mu}\geq 0.05$, with the $p_{\mu}$-value computed as follows
\beq
\label{eq3.25}
p_{\mu}=\int^{\infty}_{q_{\mu ,\mathrm{obs}}}f(q_{\mu}|\mu)\,dq_{\mu},
\eeq 
%%%%%%%%%%%%%%%%%%%%%%%%%%%%%%%%%%%
\renewcommand{\arraystretch}{1.4}
\newcolumntype{D}[1]{>{\centering\arraybackslash}p{#1}}
\begin{table}[H]
\caption{Estimates ($\hat{\mu}$) of the parameter $\mu$ obtained from the minimization of the statistic $\chi^2(\mu)$ defined in eq.~(\ref{eq3.23}). The displayed results correspond to the limiting cases $\Omega = 4\pi~\mathrm{sr}$ and $\Omega = 2\pi~\mathrm{sr}$.}
\label{table3.1}
\begin{center}
\begin{tabular}{|D{2cm}||D{1.9cm}|D{1.9cm}|}
%\begin{tabular}{|c|r||r|c||r|c||r|c|}
\hhline{|===|}
%\hhline{|--------|}
\multirow{2}{*}{$m_{\chi}~(\mathrm{GeV})$} & \multicolumn{2}{c|}{$\hat{\mu} \,(=|\lambda^1_1|^2+|\lambda^1_2|^2)$} \\[0.6mm] \cline{2-3}
& $\Omega = 4\pi~\mathrm{sr}$~~&$\Omega = 2\pi~\mathrm{sr}$\\
\hhline{|===|} 
%\hhline{|--------|}
$500$ & $-0.082$~~~ & $-0.163\,$~~~\\[0.6mm]
\hline
$600$ & $-0.059$~~~ & $\,-0.117$~~~ \\[0.6mm]
\hline
$700$ & $0.100$ & $0.199$  \\[0.6mm]
\hline
$800$ & $0.466$ & $0.931$  \\[0.6mm]
\hline
$900$ & $1.091$ & $2.182$  \\[0.6mm]
\hline
$1000$ & $1.952$ & $3.905$  \\[0.6mm]
\hline
$1100$ & $2.874$ & $5.749$  \\[0.6mm]
\hline
$1200$ & $3.467$ & $6.934$  \\[0.6mm]
\hline
$1300$ & $3.116$ & $6.232$  \\[0.6mm]
\hline
$1400$ & $0.975$ & $1.951$  \\[0.6mm]
\hline
$1500$ & $-4.224$~~~ & $\,-8.448$~~~  \\[0.6mm]
\hhline{|===|}
%\hhline{|--------|}
\end{tabular}
\end{center} 
\end{table}
%%%%%%%%%%%%%%%%%%%%%%%%%%%%%%%%%%%%%%%
\noindent
where $q_{\mu ,\mathrm{obs}}$ is the value of $q_{\mu}$ obtained from the data and $f(q_{\mu}|\mu)$ is the probability density function (pdf) of $q_{\mu}$ assuming the data correspond to the value $\mu$. In the $|\lambda^1_1|$-$|\lambda^1_2|$ plane the $95\%\,\mathrm{CL}$ contour is simply a circle of radius $\sqrt{\mu}$. For this reason, we list in table \ref{table3.2} the $95\%\,\mathrm{CL}$ upper limits on the quantity $\sqrt{\mu}$ rather than $\mu$.\par
\begin{center}
\begin{figure}[ht]
\centering
\includegraphics[scale=0.46]{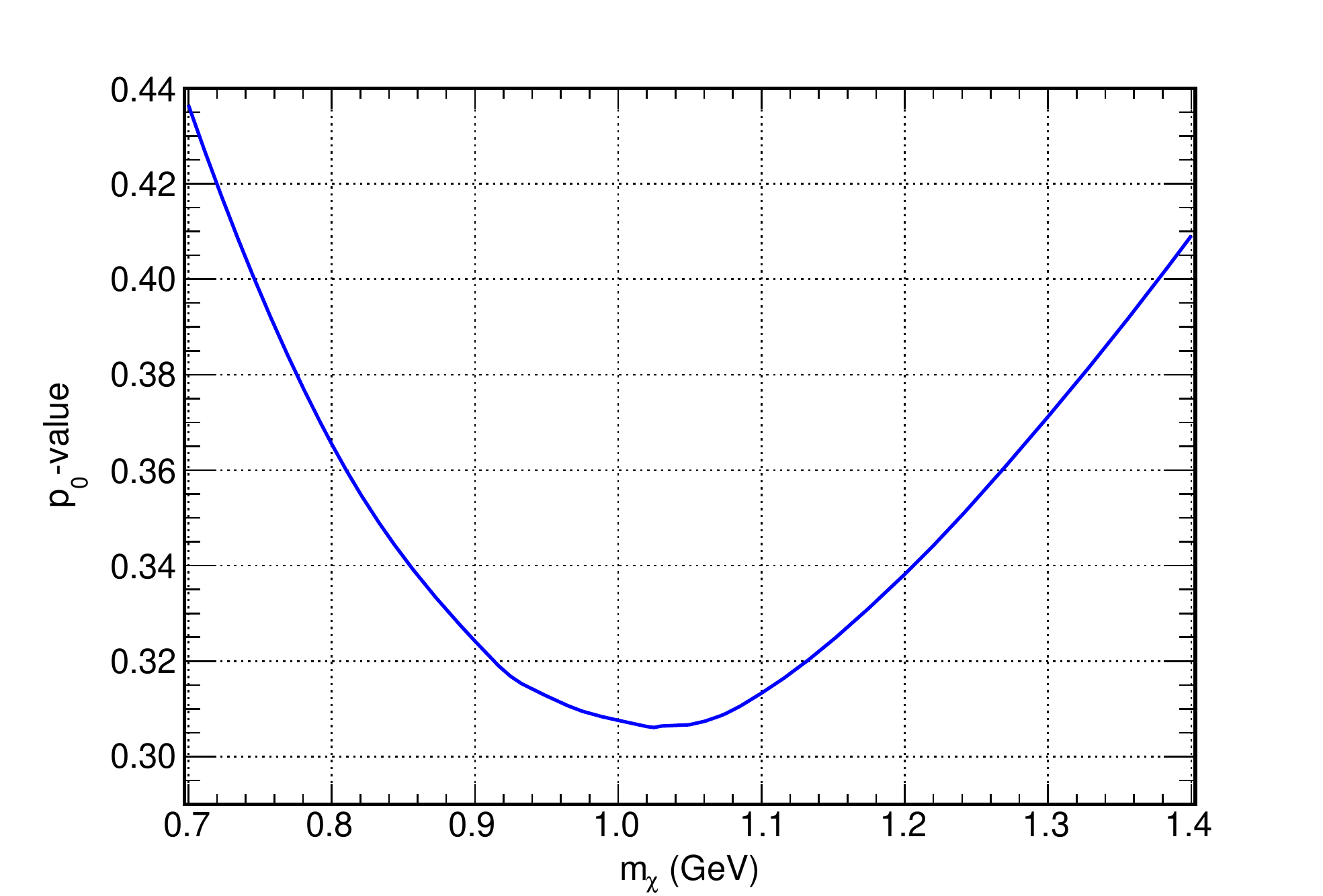}
\caption{$p_0$-value defined in eq.~(\ref{eq3.26}) as a function of the LQ mass.}
\label{fig3.5}
\end{figure}
\end{center}
\begin{center}
\begin{figure}[ht]
\centering
\hspace*{-1mm}
\subfloat{\includegraphics[scale=0.43]{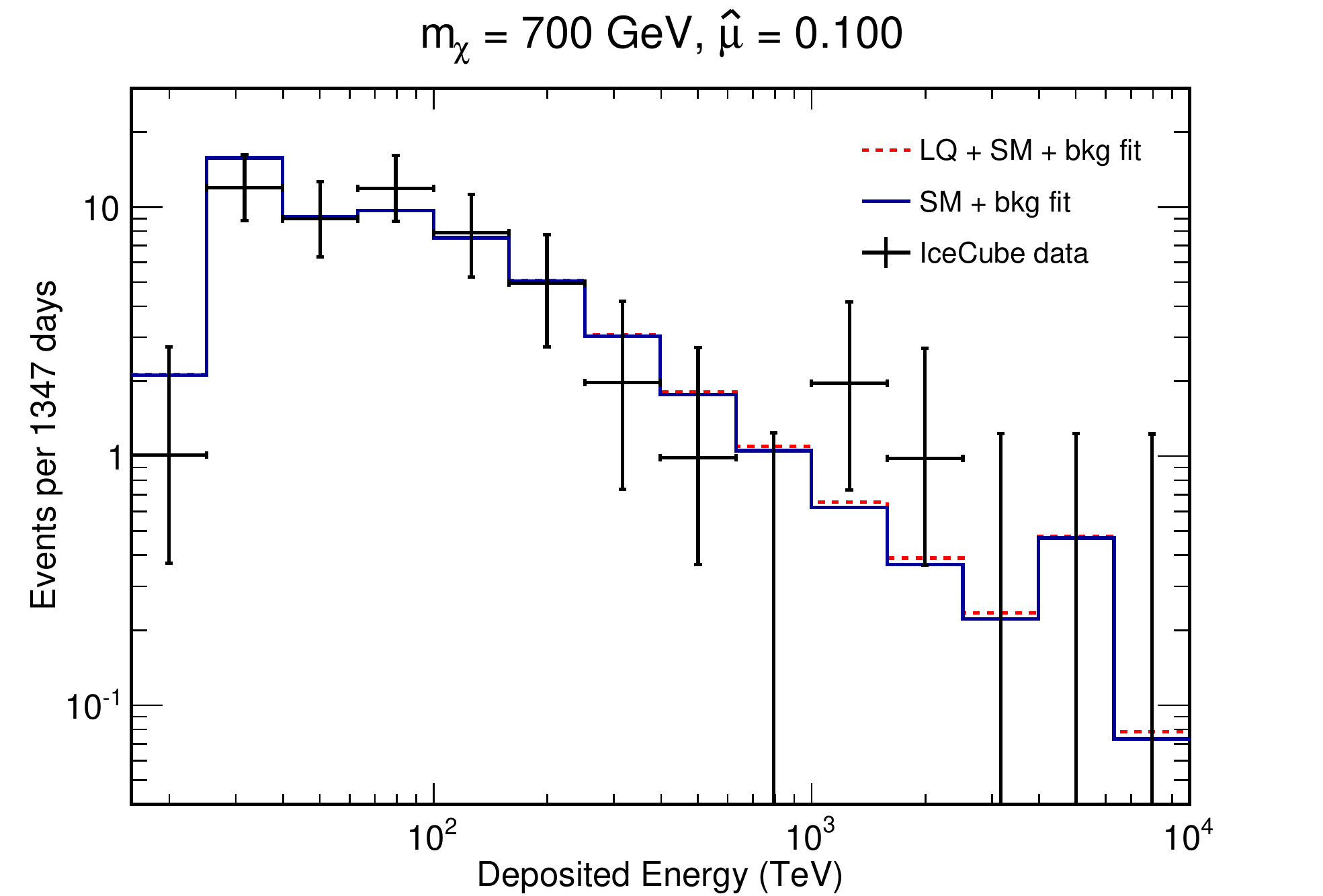}}
\hspace*{1mm}
%\hspace*{-0.03\textwidth}
%\label{fig1a}}
\subfloat{\includegraphics[scale=0.43]{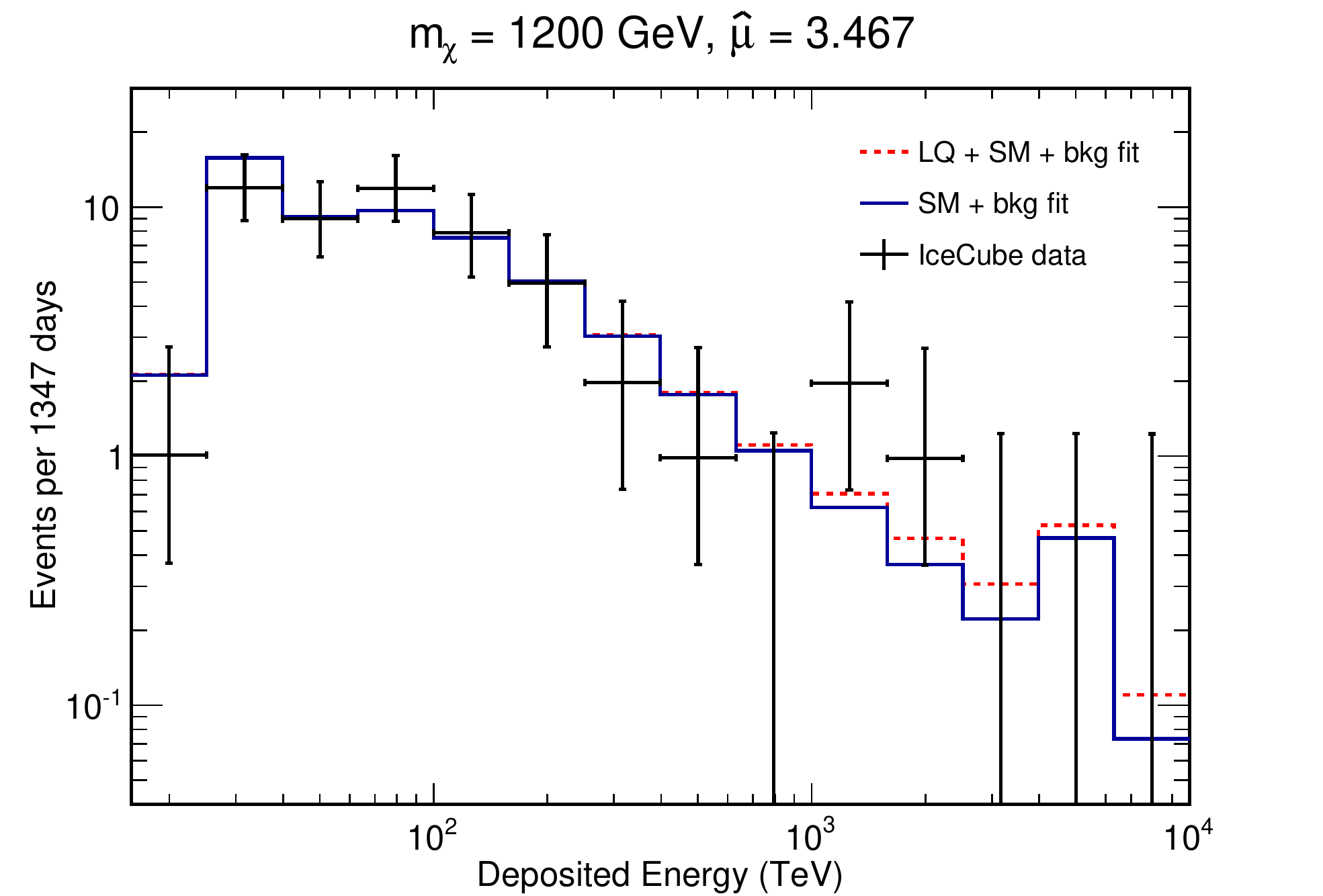}}\\[4mm]
%\hspace*{1.2cm}
\subfloat{\includegraphics[scale=0.43]{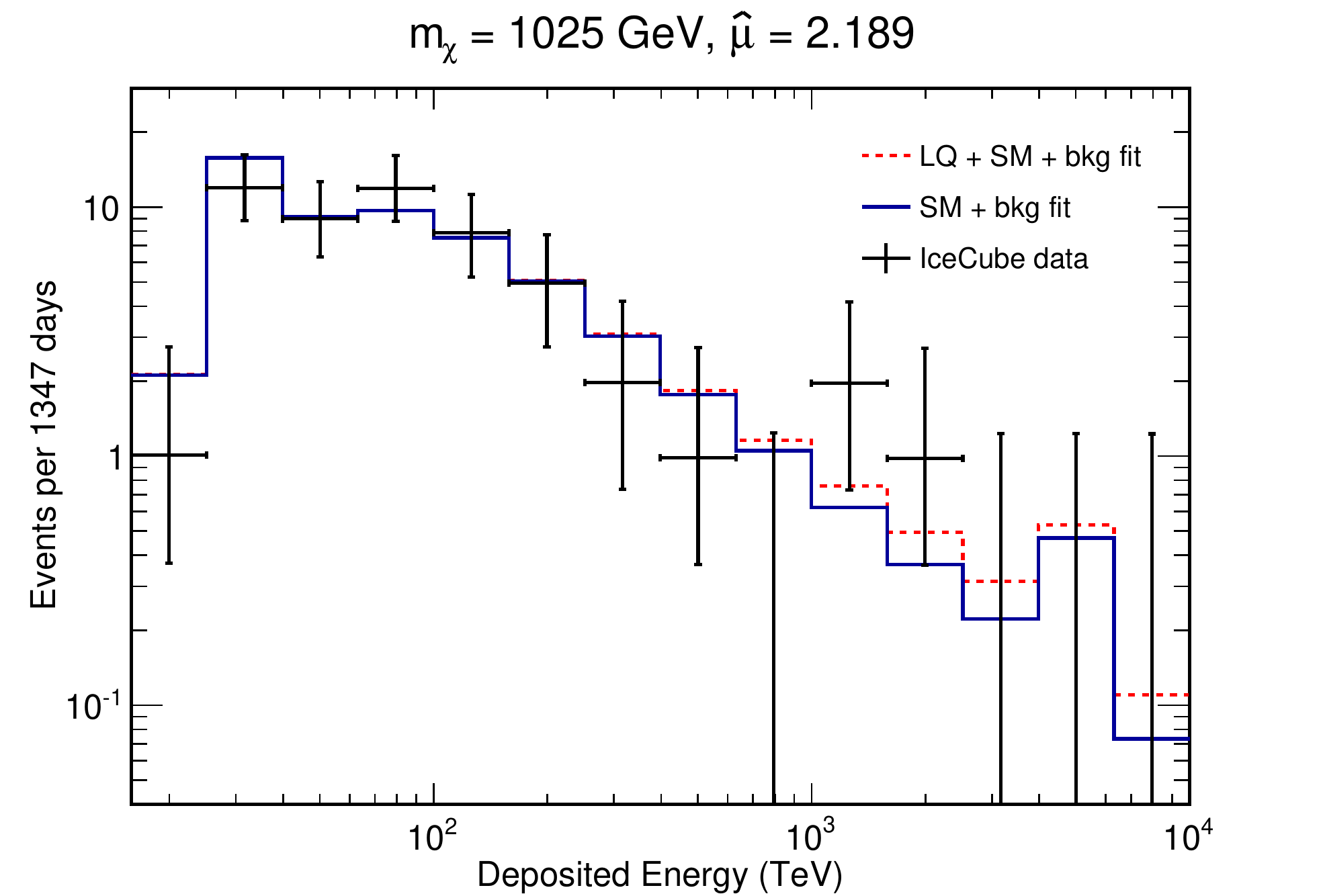}}
%\vspace*{1mm}
\caption{Total number of events observed at the IceCube along with the predictions from the SM + background (full line) and the SM + background + LQ contribution (dashed line) for the best fit case ($m_{\chi}=1025\,\mathrm{GeV}$). The fits for $m_{\chi}=700\,\mathrm{GeV}$, $1200\,\mathrm{GeV}$ are included for comparison purposes. The IceCube data as well as the SM + background fit were taken from ref.~\citep{Aartsen:2015zva}.}
\label{fig3.6}
\end{figure}
\end{center}

In order to further specify the improvement in the fit obtained by adding the LQ contribution, we quantify the level of disagreement between the data and the hypothesis $\mu=0$. For this purpose, we use the statistic test $q_0 = -2\mathrm{ln}(\lambda(0))$ for $\hat{\mu}\geq 0$, and compute the respective $p$-value as
\beq
\label{eq3.26}
p_0=\int^{\infty}_{q_{0,\mathrm{obs}}}f(q_0|0)\,dq_0,
\eeq
where $f(q_0|0)$ is the pdf of $q_0$ assuming the SM + background hypothesis ($\mu=0$). We note that the data is consider to show lack of agreement with the hypothesis $\mu = 0$ only if $\hat{\mu}>0$. Thus, we apply this test only to the LQ masses for which a physical value of $\hat{\mu}$ was obtained.
%%%%%%%%%%%%%%%%%%%%%%%%%%%%%%%%%%%%%%%%%
\begin{table}[H]
\caption{$95\%\,\mathrm{CL}$ upper limits on $\sqrt{\mu}$ obtained from eqs.~(\ref{eq3.24}) and (\ref{eq3.25}). The displayed results correspond to the limiting cases $\Omega = 4\pi~\mathrm{sr}$ and $\Omega = 2\pi~\mathrm{sr}$.}
\label{table3.2}
\begin{center}
\begin{tabular}{|D{2cm}||D{1.9cm}|D{1.9cm}|}
%\begin{tabular}{|c|r||r|c||r|c||r|c|}
\hhline{|===|}
%\hhline{|--------|}
\multirow{2}{*}{$m_{\chi}~(\mathrm{GeV})$} & \multicolumn{2}{c|}{$95\%\,\mathrm{CL} $ upper limit on $\sqrt{\mu}$} \\[0.6mm] \cline{2-3}
& $\Omega = 4\pi~\mathrm{sr}$~~&$\Omega = 2\pi~\mathrm{sr}$\\
\hhline{|===|} 
%\hhline{|--------|}
$500$ & $0.687$ & $0.971$\\[0.6mm]
\hline
$600$ & $1.074$ & $1.519$ \\[0.6mm]
\hline
$700$ & $1.572$ & $2.224$  \\[0.6mm]
\hline
$800$ & $2.181$ & $3.085$  \\[0.6mm]
\hline
$900$ & $2.937$ & $4.154$  \\[0.6mm]
\hline
$1000$ & $3.781$ & $5.345$  \\[0.6mm]
\hline
$1100$ & $4.774$ & $6.752$  \\[0.6mm]
\hline
$1200$ & $5.856$ & $8.281$  \\[0.6mm]
\hline
$1300$ & $7.015$ & $9.921$  \\[0.6mm]
\hline
$1400$ & $8.337$ & $11.790$  \\[0.6mm]
\hline
$1500$ & $9.599$ & $13.575$  \\[0.6mm]
\hhline{|===|}
%\hhline{|--------|}
\end{tabular}
\end{center} 
\end{table}
%%%%%%%%%%%%%%%%%%%%%%%%%%%%%%%%%%%%%%%%%
\vspace*{-4mm}
In figure~\ref{fig3.5} we show the $p_0$-value as a function of the LQ mass in the range $[700~\mathrm{GeV},1.4~\mathrm{TeV}]$. We see that the hypothesis $\mu = 0$ cannot be rejected conclusively in any of the considered cases. For $m_{\chi}=700\,\mathrm{GeV}$, the level of disagreement between data and the SM+backgound hypothesis is such that the latter could be rejected at a confidence level of $56\%$. This confidence level increases with the LQ mass and attains its maximum (minimum $p_0$), given by $\sim\!69.5\%$, at $m_{\chi}\simeq 1025\,\mathrm{GeV}$. In the lower panel of figure~\ref{fig3.6}, we show the total number of events observed at Icecube along with the predictions from the SM + background component and when the LQ contribution corresponding to $m_{\chi}=1025\,\mathrm{GeV}$ is added on top of it. For masses deacreasing from $1025\,\mathrm{GeV}$ to $700\,\mathrm{GeV}$, the $p_0$-value increases, indicating that the fit worsen (see the upper left panel of figure~\ref{fig3.6}). Since the LQ contribution of the smaller masses affects mainly the bins of deposited energy where the majority of the events appear, the corresponding $\hat{\mu}$ is forced to small values which leads to a negligible impact on the two bins that exhibit a weaker agreement with the SM + background explanation. As long as the LQ mass increases, their contributions become maximum at the region around these two bins, improving the fit with higher values of $\hat{\mu}$ (see table \ref{table3.1}). The $p_0$-value also increases for masses higher than the best fit value because the maximum of the respective LQ contributions moves away from the two bins between $2$-$3\,\mathrm{PeV}$ and start to affect the most energetic bins in which no events have been observed (see, for example, the upper right panel of figure~\ref{fig3.6}). This leads to worse fits and for $m_{\chi}>1200\,\mathrm{GeV}$ pushes the $\hat{\mu}$ again towards smaller values (see table \ref{table3.1}). 
%%%%
%%%%
\vspace*{-2cm}
%%%%
\vspace*{-1cm}
%%%%%%%%%%%%%%%%%%%%%%%%%
%%%%%%%%%%%%%%%%%%%%%%%%%%%%%%%%%%%%%%%%%%%%%%%%%%%%%%%%%%%%%%%%%%%%%%%%% 
\section{LHC Constraints}
\label{sec4}
\begin{center}
\begin{figure}[ht]
\centering
\hspace*{1mm}
\subfloat{\includegraphics[scale=0.39]{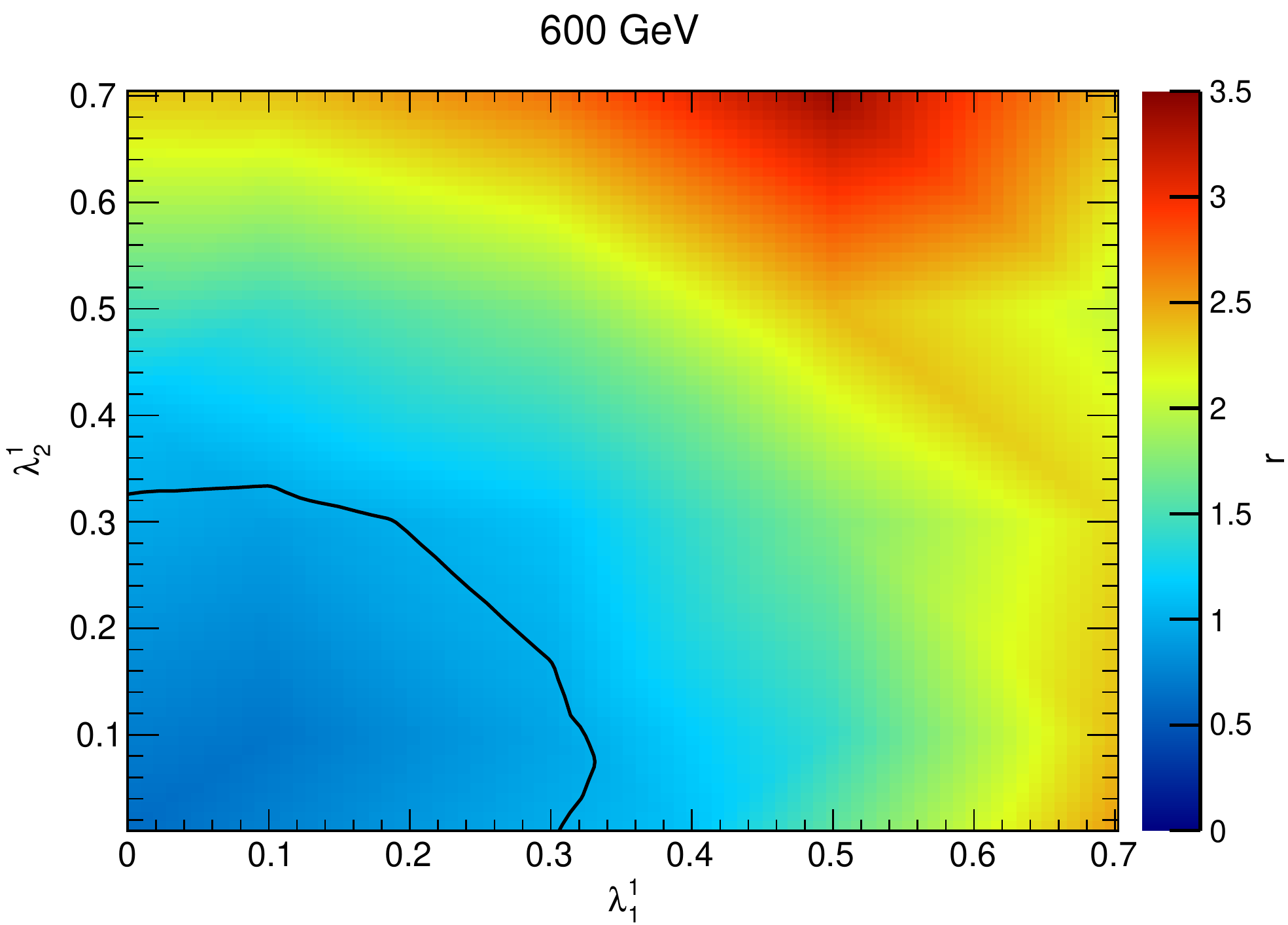}}
\hspace*{2.2mm}
%\hspace*{-0.03\textwidth}
%\label{fig1a}}
\subfloat{\includegraphics[scale=0.39]{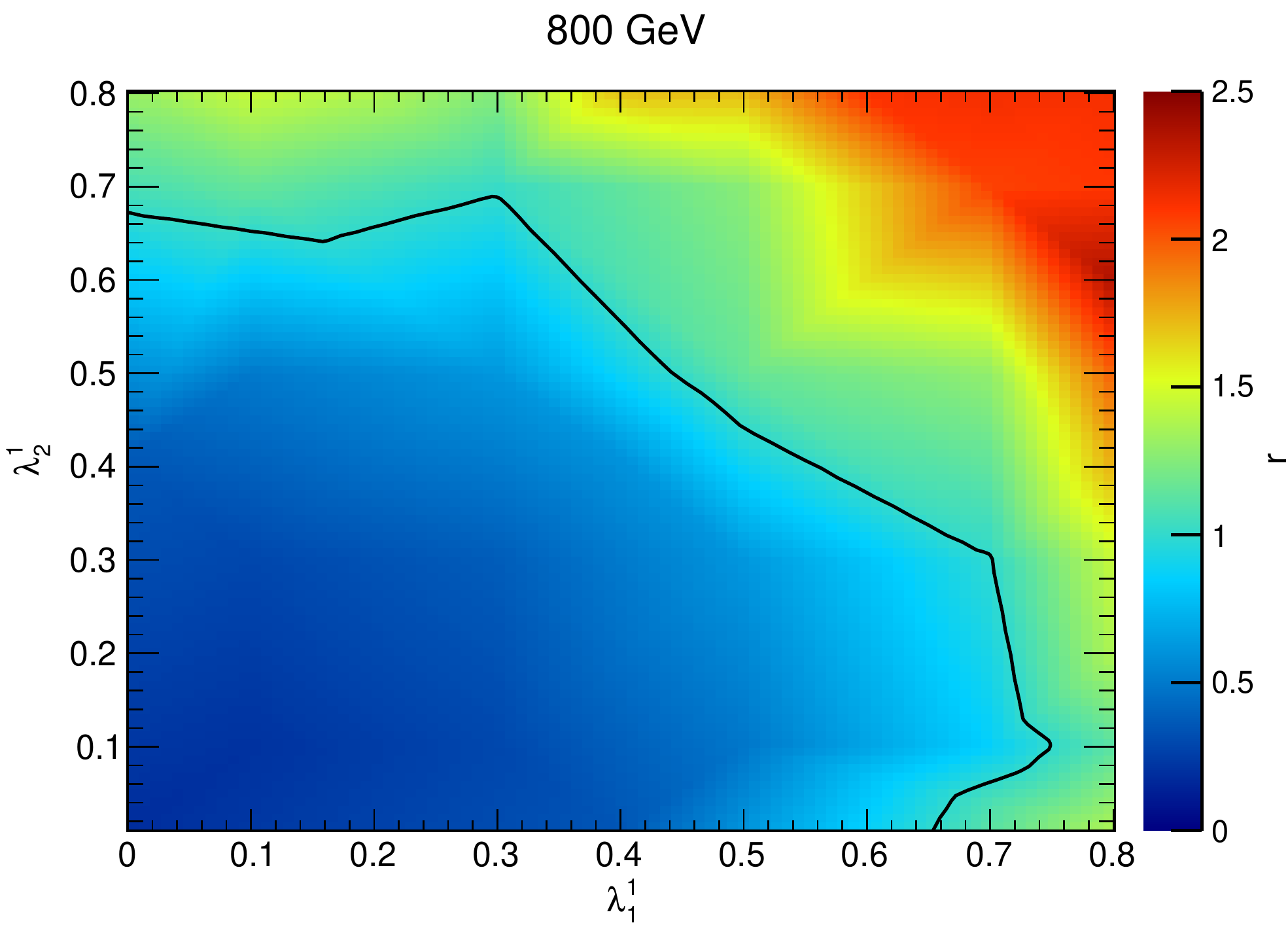}}\\[3.5mm]
\hspace*{1.1mm}
\subfloat{\includegraphics[scale=0.395]{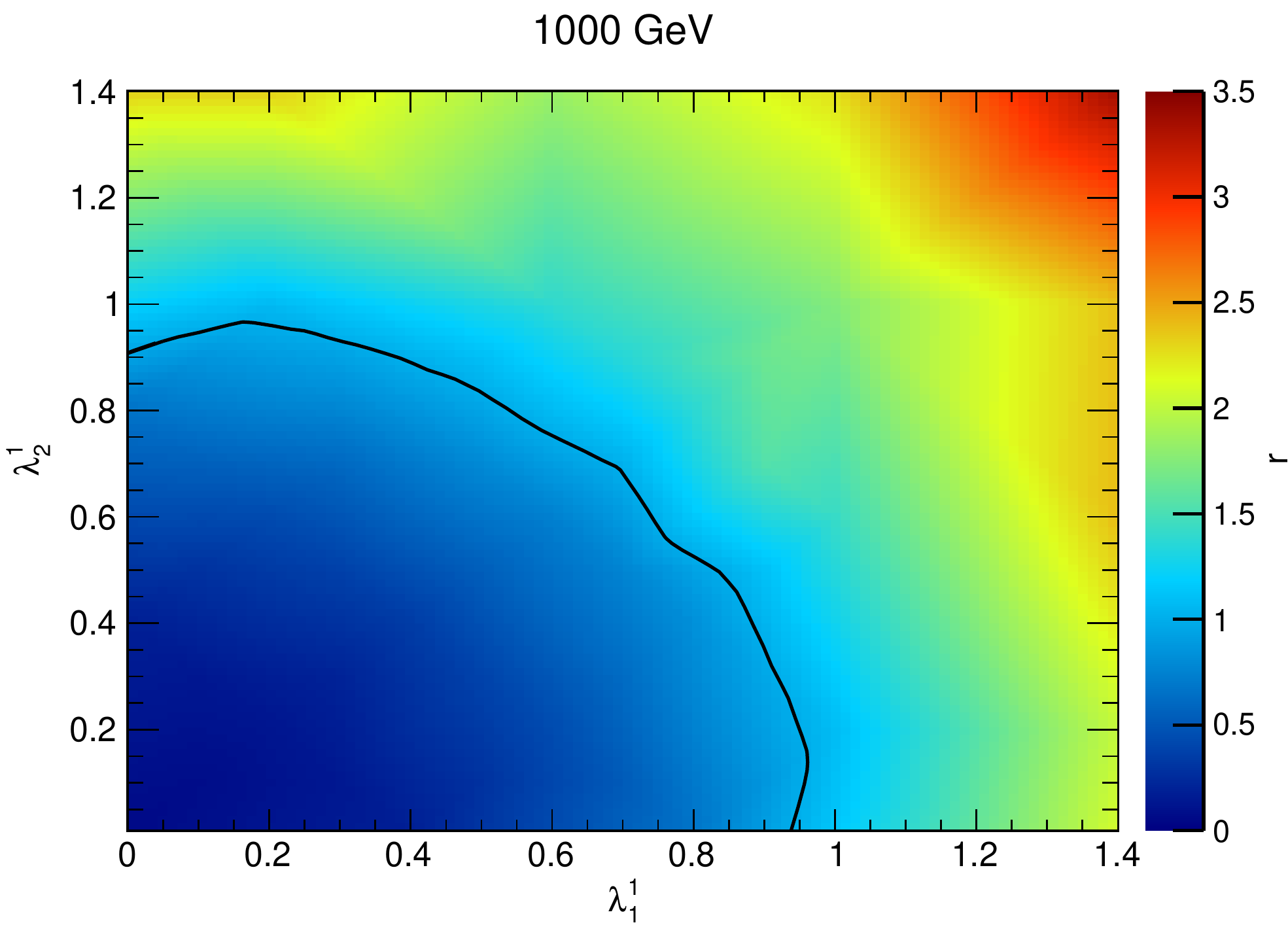}}
\hspace*{1.4mm}
\subfloat{\includegraphics[scale=0.39]{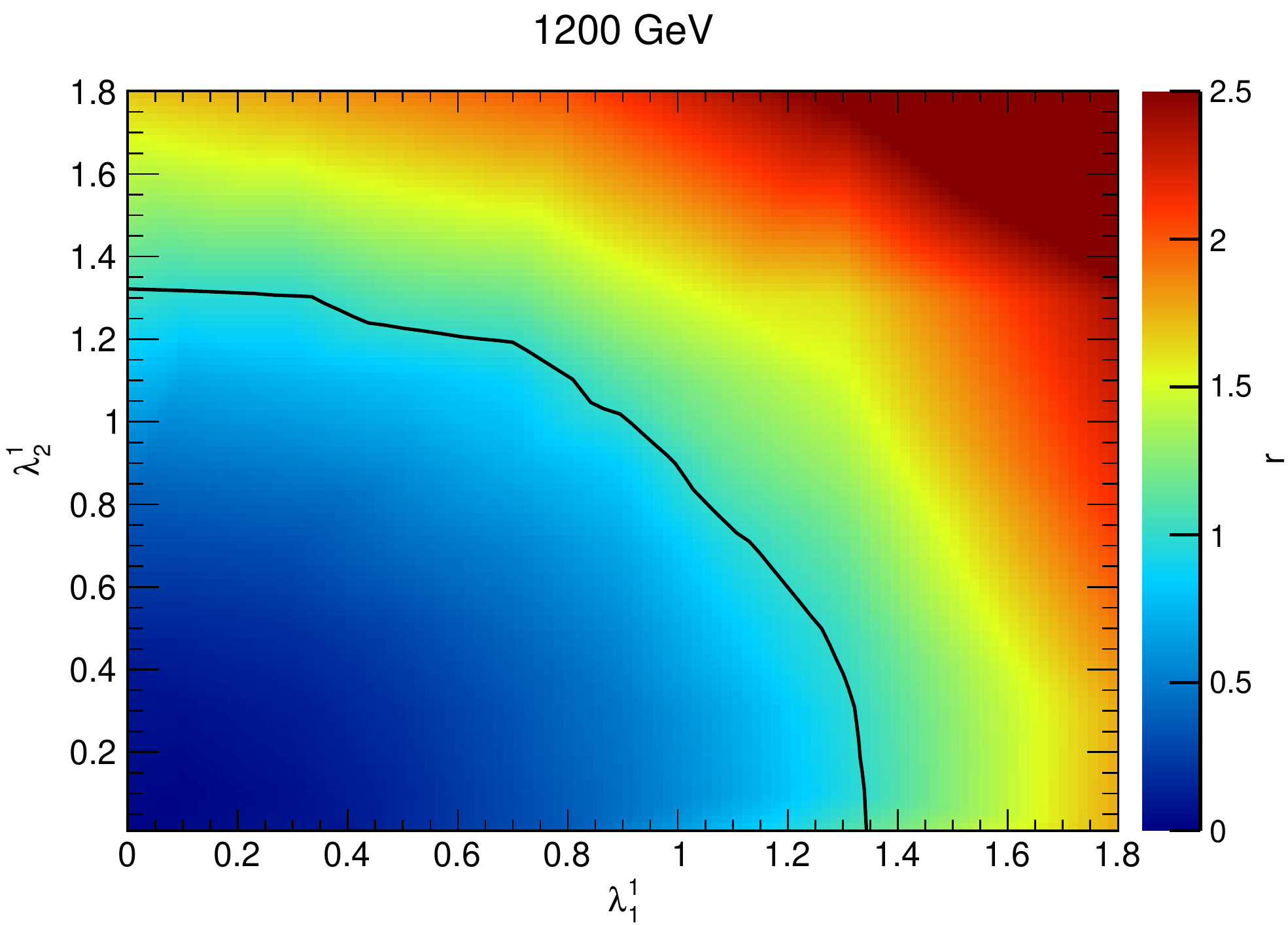}}
\vspace*{1mm}
\caption{Allowed region ($r<1$) in the $\lambda^{1}_{1}$-$\lambda^{1}_{2}$ plane with $\lambda^i_j=0$ if $i\neq 1$ and/or $j=3$. From top left to bottom right: $m_{\chi}=600\,$-$\,1200$ GeV in 200 GeV increments. The results were obtained using all final states denoted by topologies (a) and (b).}
\label{fig4.1}
\end{figure}
\end{center}

In this section we discuss the most up-to-date LHC constraints on our colored electroweak-triplet scalar. We begin our discussion by filtering the model through the latest $8$ TeV data. Our framework leads to five distinct final state topologies that we classify as follows
\begin{itemize}
\item 2 jets + MET (a)
\item 1 jet + MET (a)
\item 2 jets + 2 leptons (b)
\item 1 lepton + 1 jet + MET (b)
\item 2 leptons + 1 jet (b)
\end{itemize}
We simulate the (a) and (b) topologies separately using MadGraph 5~\cite{Alwall:2014hca}. We implement PYTHIA~\cite{Sjostrand:2006za} for the parton shower and hadronization and the detector simulation is carried out using Delphes 3~\cite{deFavereau:2013fsa}. We simulate the two topologies in a separate manner since the lepton misidentification rate is very small and events with final states containing only jets and missing energy will not significantly contribute to final state topologies containing leptons. In fact, the electron fake rate can be anywhere between $10^{-4}$ and $10^{-5}$~\cite{Giordano:2013xba} while a recent study finds a muon fake rate of $2\times10^{-5}$~\cite{1505.01934}. The events are generated for masses in the range $600 < m_{\chi} <1200$ GeV, for different combinations of the couplings assuming that $\lambda^{i}_{j}=0$ if $i\neq 1$ and/or $j=3$. In order to set bounds on the parameter space of the model, we use the latest CheckMATE validated analyses~\cite{CheckMate}:
\begin{itemize}
\item ATLAS search for squarks and gluinos with jets and missing momentum~\cite{1405.7875},
\item ATLAS search for third generation squarks via charm quarks or compressed supersymmetric scenarios~\cite{1407.0608},
\item ATLAS search for new phenomena with high energetic jets and large missing transverse momentum~\cite{1502.01518}
\end{itemize}
for the (a) topologies, and 
\begin{itemize}
\item ATLAS search for direct top-squark pair production in final states with two leptons~\cite{1403.4853},
\item ATLAS search for top squark pair production with one isolated lepton and missing transverse momentum~\cite{1407.0583},
\item ATLAS search for supersymmetry in events containing a same-sign dilepton pair, jets and large missing transverse momentum~\cite{1503.03290}, 
\item ATLAS search for direct slepton and chargino production in final states with two opposite-sign leptons, missing energy and no jets~\cite{atlas-conf-2013-049}
\end{itemize}
for the (b) topologies. In figure~\ref{fig4.1} we show results for LQ masses in the range $600$-$1200$ GeV in the $\lambda^{1}_{1}$-$\lambda^{1}_{2}$ plane after applying all of the $8$ TeV LHC results listed above. We compare our results to the $95\%$ upper confidence limits on the number of signal events using the variable $r$ defined in~\cite{CheckMate} given by
\begin{equation}
r=\frac{S-1.96\cdot\Delta S}{S^{95}_{exp}},
\end{equation}
where the numerator parametrizes the $95\%$ lower limit on the number of signal events determined by CheckMATE and the denominator the $95\%$ experimental limit on the number of signal events. Regions of parameter space are excluded if $r\ge1$. In figure~\ref{fig4.1} we depict, for all LQ masses, the $r=1$ contour with a black solid line. We do not show results for masses below $600$ GeV, since for couplings $\lambda^{1}_{j}>0.1$, which is the case for the simulations performed in this work, these masses are not allowed by current experimental constraints. 
We also note that ATLAS and CMS have dedicated searches for first and second generation LQs with the $8$ TeV~\cite{Aad:2015caa,Khachatryan:2015vaa} and $13$ TeV~\cite{Aaboud:2016qeg,CMS:2016qhm} data sets, with a slight improvement on the limits with the latter. The searches target LQ pair production. The CMS collaboration focuses primarily on the second generation and place limits of 1165 and 960 GeV for LQ branching fractions of 0.5 and 1 respectively using 2.7 fb$^{-1}$ of data, while the ATLAS collaboration places limits of 1100 and 1050 GeV for first and second generation LQs respectively when the branching ratio is $100\%$ to a lepton and a quark. In addition, the ATLAS collaboration obtains limits varying the branching ratio into electrons and muons which are shown in figure 7 of~\cite{Aaboud:2016qeg}.

In order to apply these LQ dedicated searches to our model we follow a conservative approach since the only component of $\chi$ that decays purely to a charged lepton and a quark is $\chi_{3}$; hence limits on the mass of $\chi_{1}$ and $\chi_{2}$ will turn out to be much weaker. However, since we are assuming mass degeneracy to avoid tensions with electroweak precision data (EWPD)~\cite{Dorsner:2016wpm,EWPD}, the limits apply across the components of $\chi$. In addition, since we are assuming that $\chi$ primarily couples the first family of quarks to electrons and muons, the decay width must be saturated with these two decay modes. As a consequence, the constraints given in figure 7 of ~\cite{Aaboud:2016qeg} only imply that our LQ must lie above $900$ GeV. Therefore, our model is basically unconstrained by the LQ searches and the limits derived from the more general searches described above dominate.
%
%%%%%%%%%%%%%%%%%%%%%%%%%%%%%%%%%%%%%%%%%%%%%%%%%%%%%%%%%
\section{Low energy physics observables}
\label{sec5}
The renormalizable interactions introduced in eq.~(\ref{eq2.3}) can lead to rare flavor changing and CP violating processes both at tree-level and at the one-loop level. Our working assumption is that $\chi$ couples primarily the first family of quarks to the electron and the muon and helps us to avoid the most stringent bounds arising from tree level semi-leptonic and leptonic meson decays as well as semi-leptonic $\tau$ decays. However, our LQ can yield new contributions to muon rare decays such as $\mu\to e\gamma$, the magnetic dipole moment of the muon, and atomic parity violation measurements. We discuss these constraints below.
\subsection{$\mu\to e\gamma$ and $(g-2)_{\mu}$}
\label{sec5.1}
\begin{center}
\begin{figure}[ht]
\centering
%\hspace*{-0.4cm}
\subfloat{\includegraphics[scale=0.43]{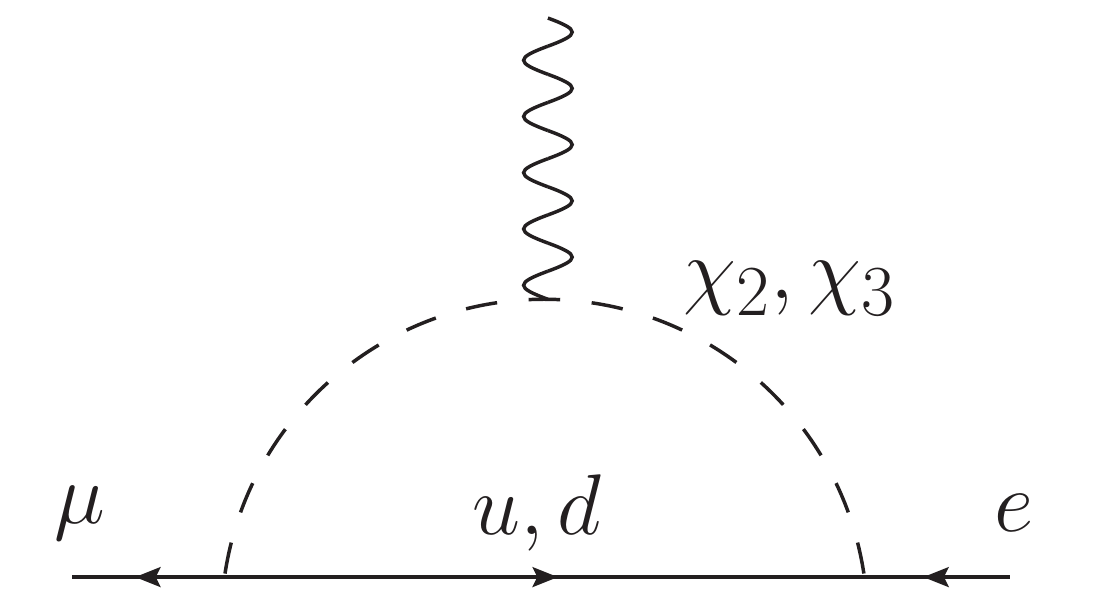}}
\hspace*{0.1\textwidth}
%\label{fig1a}}
\subfloat{\includegraphics[scale=0.43]{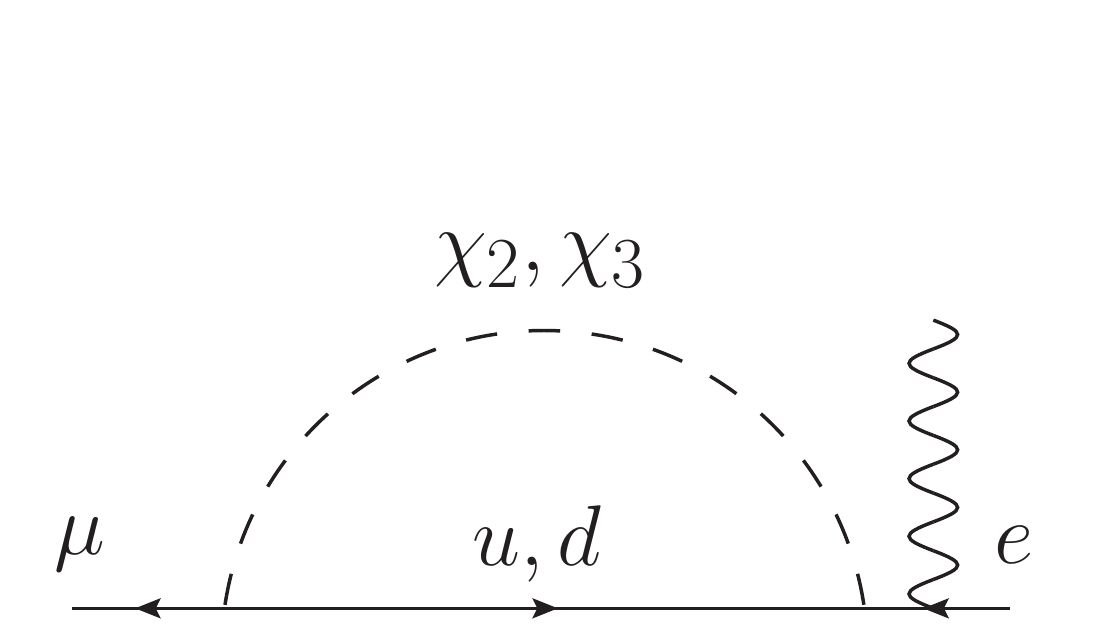}}\\[6mm]
%\label{fig1b}}
\subfloat{\includegraphics[scale=0.43]{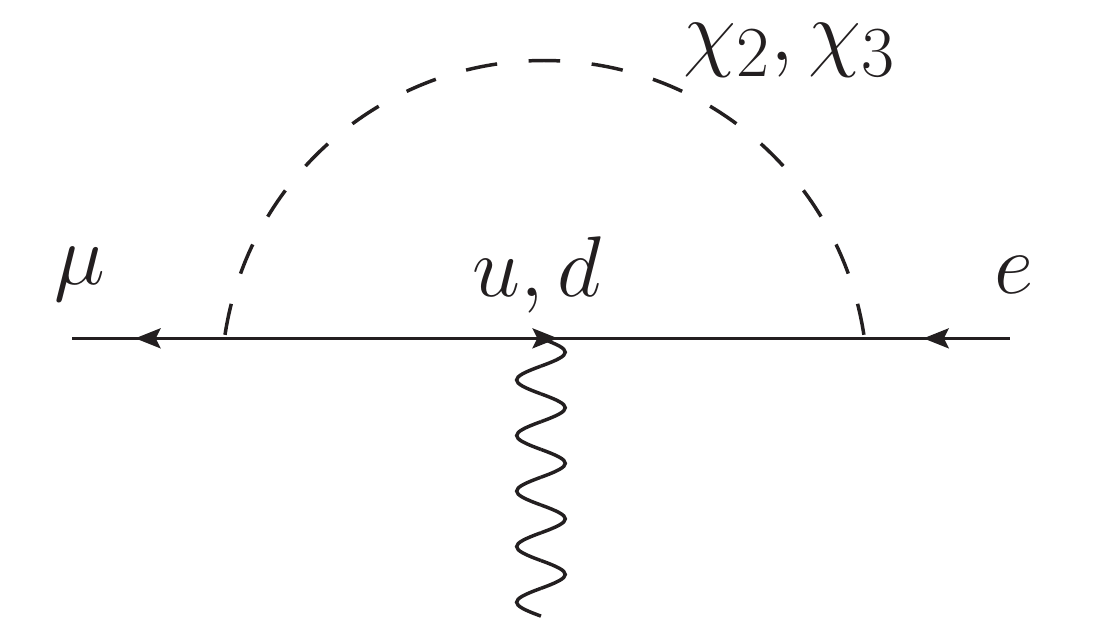}}
\hspace*{0.1\textwidth}
\subfloat{\includegraphics[scale=0.43]{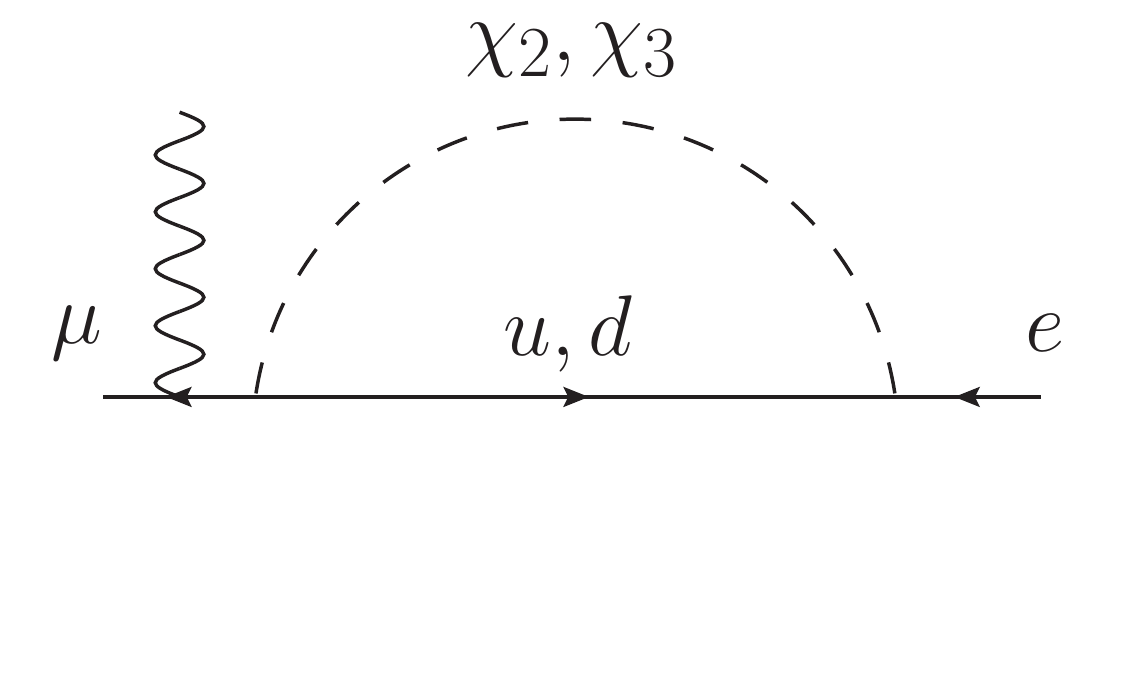}}
\caption{1-loop diagrams for $\mu \to e \gamma$ decays. The arrows indicate fermion charge flow. Main contribution comes from the diagrams in the first column whereas those in the second one are needed to enforce gauge invariance.}
\label{fig:mueg}
\end{figure}
\end{center}

 Our LQ, a colored electroweak-triplet scalar, can give rise to lepton flavor violating decays such as $\mu \to e \gamma$ as well as a contribution to the muon anomalous magnetic moment, $a_\mu$. Both contributions come in at the 1-loop level. The Feynman diagrams for the $\mu \to e \gamma $ decay process are  depicted in figure~\ref{fig:mueg}.\par
To place constraints in our model we follow the conventions used in~\cite{Dorsner:2016wpm} where the relevant parts of eq.~(\ref{eq2.3}) contributing to the $\mu\to e\gamma$ decay and the muon's $(g-2)$ can be expressed as 
%%%%%%%%%%%%%%%%%%%%%%%%%%%%%%%%
\begin{equation}
{\cal L}\supset \lambda^{i}_{j}\,\bar{d}^{\,c}_{\,iL}\,\chi^{\dagger}_{3}\,e_{jL}-(1/\sqrt{2})(V^{T}\lambda)^{i}_{j}\,\xbar{u}^{\,c}_{\,iL}\,\chi^{\dagger}_{2}\,e_{jL}+~\text{h.c.},
\end{equation}
where $V$ is the Cabibbo-Kobayashi-Maskawa mixing matrix. The above expression was obtained by starting from a mass-ordered mass eigenstate basis for the down type quarks and charged leptons and applying the following transformations: $u_{iL}\to (V^{\dagger})_{ik}\,u_{kL}$, $d_{iL}\to d_{iL}$, and $e_{jL}\to e_{jL}$. Since $V_{12}$ is roughly $20\%$ of $V_{11}$, we will assume that the coupling of both the muon and the electron to down- and up-type quarks is the same. With this working assumption, and using the following effective Lagrangian for the $\mu\to e\gamma$ decay
\begin{equation}
\mathcal{L} = A\,\bar{e}\,i\sigma^{\mu\nu}(1+\gamma^{5})\,\mu\, F_{\mu\nu},
\end{equation}
%%%%%%%%%%%%%%%%%%%%%%%%%%%%%%%%%%%% 
%%%%%%%%%%%%%%%%%%%%%%%%%%%%%%%%%%%%%%%
the decay width is given by
\begin{equation}
\label{eq:muegwidth}
\Gamma (\mu\to e \gamma) = \frac{|A|^2 m_\mu ^3}{16\pi}.
\end{equation}
By assuming non-negligible couplings of the electron and muon to the first family only, a standard calculation yields the following expression for $A$
\begin{equation}
\label{eq:A}
A=\frac{3e}{64 \pi^2}\left( \frac{m_\mu}{m_\chi ^2}\right)\lambda^1_2 \lambda^1_1,
\end{equation}
where a common mass, $m_{\chi}$, have been used for both $\chi_1$ and $\chi_2$, and terms of ${\cal O}\left(m_{u(d)}/m_{\chi}\right)$ have been neglected.
%
%\begin{equation}
%\begin{split}
%\label{eq:A}
%A&=\frac{e m_\mu}{64\pi^2 m_\chi ^2}\sum_{i=t,c,u}\frac{\lambda_\mu ^i \lambda_e ^i}{(1-a_i)^3}\left[ 3-10 a_i -5 a_i^2 +\frac{2a_i -16a_i ^2}{1-a_i}\ln a_i\right] \\
%&\underset{a_i\to 0}{\longrightarrow} \frac{3e}{64 \pi^2}\left( \frac{m_\mu}{m_\chi ^2}\right)\lambda_\mu ^1 \lambda_e ^1 ,
%\end{split}
%\end{equation}
%where in the second line we have assumed non-negligible couplings of the electron and muon to the first quark family only. In the above equation we have assumed a common mass, $m_\chi$, for both $\chi_2$ and $\chi_3$ and defined $a_i=m_i^2/M_\chi ^2$. We have also neglected terms of ${\cal O}\left((m_{t(b)}/M)^2\right)$. In the limit where the LQ only couples the electron and muon to the first quark family, the branching ratio is given by
The branching ratio is then given by
\begin{equation}
Br(\mu\to e \gamma) =1.8\left( \frac{ \mathrm{TeV}}{m_\chi}\right)^4 \times 10^{-6}|\lambda^{1}_{2}\lambda^{1}_{1}|^2.\label{eq:Brmutoeg1}
\end{equation}
In order to extract an upper bound on the value of $\lambda^{1}_{2}\lambda^{1}_{1}$ we use the current $\mu\to e\gamma$ experimental bound, $Br\left(\mu\to e\gamma\right)\le4.2\times10^{-13}$, published by the MEG collaboration~\cite{TheMEG:2016wtm} to arrive at
\begin{equation}
|\lambda^{1}_{2}||\lambda^{1}_{1}|\le4.83\times10^{-4}\left(\frac{m_{\chi}}{\text{TeV}}\right)^{2}.
\end{equation}

In the same way, loops of LQs can modify the anomalous magnetic moments of leptons, $a_{l}$. The effective Lagrangian parameterizing modifications to $a_{l}$ can be written as 
\begin{equation}
{\cal L}_{a_{l}}=e\cdot\bar{l}\left(\frac{a_{l}}{4m_{l}}\sigma_{\mu\nu}F^{\mu\nu}\right)l.
\end{equation}
The contribution to $a_{\mu}$ in the $m_{q}/m_{\chi}\to 0$ limit from $\chi_{2}$ and $\chi_{3}$ is given by~\cite{Dorsner:2016wpm}
\begin{equation}
a_{\mu}\approx\frac{9}{32\pi^{2}}\frac{m^{2}_{\mu}}{m^{2}_{\chi}}(1+2\sqrt{2})|\lambda^{1}_{2}|^2.
\end{equation}
The most precise experimental result on $(g-2)_{\mu}$ was obtained by the E821 experiment carried out at BNL~\cite{Bennett:2004pv,Bennett:2006fi}. The deviation from the SM value is given by $\delta a_{\mu}=(2.8\pm0.9)\times10^{-9}$ where the SM value is given by $a^{\text{SM}}_{\mu}=1.16591803(70)\times10^{-3}$~\cite{Agashe:2014kda}. Using this result we can directly constrain the value of $\lambda^{1}_{2}$:
\begin{equation}
|\lambda^{1}_{2}|\lesssim 1.5\left(\frac{m_{\chi}}{\text{TeV}}\right).
\end{equation}

From the two constraints discussed above, one can see that one scenario of interest could lead to a very suppressed value of $\lambda^{1}_{1}$ compared to $\lambda^{1}_{2}$. In particular, for LQ masses in the TeV range, one needs $\lambda^{1}_{1}\sim 10^{-3}$ for ${\cal O}\left(1\right)$ $\lambda^{1}_{2}$ couplings. These scenarios are not unnatural if one takes into account specific flavor models where quarks transform as different non-trivial singlets of $A_{4}$~\cite{Varzielas:2015iva}. Below we will discuss how this specific scenario is also consistent with low energy precision measurements such as atomic parity violation.

%%%%%%%%%%%%%%%%%%%%%%%%%%%%%%%%%%
\subsection{Atomic Parity Violation}
\label{sec5.2}
Below the electroweak scale the parity violation such as in the Cesium 133 atom can be studied with the following effective Lagrangian
\begin{equation}
{\cal L}_{PV}=\frac{G_{F}}{\sqrt{2}}\bar{e}\gamma^{\mu}\gamma^{5}e\left( C_{1u}\bar{u}\gamma_{\mu}u+C_{1d}\bar{d}\gamma_{\mu}d\right).
\end{equation}
The SM maximally violates parity and one can calculate very precisely the values of $C_{1u}$ and $C_{1d}$ with $C^{\text{SM}}_{1u}=(-1/2+4/3\sin^{2}\theta_{W})$ and $C^{\text{SM}}_{1d}=(1/2+2/3\sin^{2}\theta_{W})$, where $\theta_{W}$ denotes the Weinberg angle of the SM. Using these values one can define a nuclear weak charge by 
\begin{equation}
\label{eq:apv}
Q_{W}(Z,N)=-2[(2Z+N)C_{1u}+(2N+Z)C_{1d}],
\end{equation}
where $Z$ and $N$ are the number of protons and neutrons respectively. For cesium, the experimentally measured value of $Q_{W}$ is $-73.20(35)$~\cite{Guena:2004sq}. Using this measurement one can extract strong constraints on the LQ couplings to the first quark family. In particular, one can parametrize the contributions arising from $\chi$ by $\delta C_{1u}$ and $\delta C_{1d}$. Given that in the SM $Q_{W}=-73.15(35)$~\cite{APVSM}, using eq.~(\ref{eq:apv}) with $C_{1u}=C^{\text{SM}}_{1u}+\delta C_{1u}$ and $C_{1d}=C^{\text{SM}}_{1d}+\delta C_{1d}$ and assuming that the coupling of the electron to the up- and down-type quarks is the same, as discussed in the previous section, one can extract the following matching contribution to $\delta C_{1u}=\delta C_{1d}=\delta C_{1}$~\cite{Dorsner:2016wpm}:
\begin{equation}
\delta C_{1}=\frac{1}{G_{F}}\frac{|\lambda^{1}_{1}|^{2}}{8m^{2}_{\chi}},
\end{equation}
where $G_{F}$ denotes the Fermi constant. With the above result and the experimentally measured value of $Q_{W}$ one has the following bound on $\lambda^{1}_{1}$:
\begin{equation}
|\lambda^{1}_{1}|\lesssim0.37\left(\frac{m_{\chi}}{1~\text{TeV}}\right),
\end{equation}
which is roughly four times stronger than the bound on $\lambda^{1}_{2}$ derived from the measurement of the muon anomalous magnetic moment. In light of this result and the bound arising from the $\mu\to e \gamma$ rare decay, our framework leans towards values of $\lambda^{1}_{1}$ which are suppressed relative to $\lambda^{1}_{2}$.
%%%%%%%%%%%%%%%%%%%%%%%%%%%%%%%%%%%%%%%%%%%%%%%%%%%%%%%%%
\section{Discussion and Concluding Remarks}
\label{sec6}
Taking into account the analysis performed in sec.~\ref{sec3}, we conclude that, in order to improve the explanation of the spectrum of UHE neutrinos observed at IceCube through the addition of a LQ triplet, higher values for the mass $m_{\chi_1}=m_{\chi_2}=m_{\chi}$ are preferred. Additionally, since the rate of events expected from the LQ component decreases with the LQ mass, large values of $\mu=|\lambda^1_1|^2+|\lambda^1_2|^2$ are also required (see table \ref{table3.1}). Specifically, under the hypotheses used in eq.(\ref{eq3.21}) and described in sec.~\ref{sec3.2}, we have found that the best fit of the four year IceCube data is achieved when the LQ mass is approximately $1025\,\mathrm{GeV}$ and the couplings are such that $\mu=2.189$ (see figure~\ref{fig3.6}). We note that this mass is allowed by the dedicated searches of LQs in the LHC at $8\,\mathrm{TeV}$ and $13\,\mathrm{TeV}$.\par 
Regarding the $95\%\,\mathrm{CL}$ limits derived from the IceCube data in table \ref{table3.2}, we see that these are considerably weaker than the constraints placed by the general searches at the LHC at $8\,\mathrm{TeV}$ listed in sec.~\ref{sec4} (see figure~\ref{fig4.1}). This is mainly due to the lack of statistics in the most energetic bins of the IceCube spectrum, where the data is not sufficiently explained by the SM expectation and the LQ contribution may become more relevant.\par
%%%%%%%%%%%%%%%%%%%%%%%%%%%%
\begin{center}
\begin{figure}[ht]
\centering
\hspace*{1mm}
\subfloat{\includegraphics[scale=0.36]{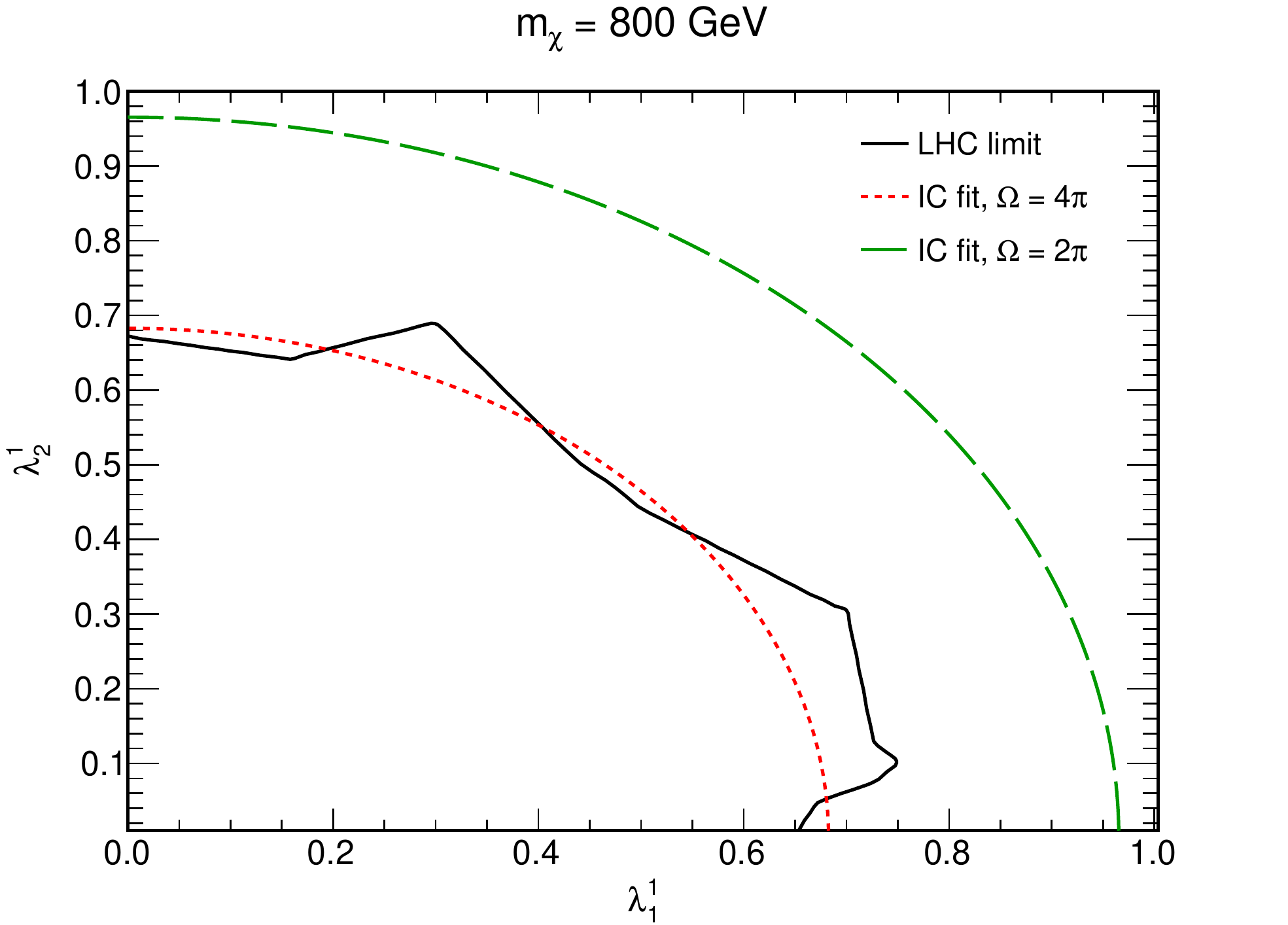}}
\hspace*{1mm}
%\hspace*{-0.03\textwidth}
%\label{fig1a}}
\subfloat{\includegraphics[scale=0.36]{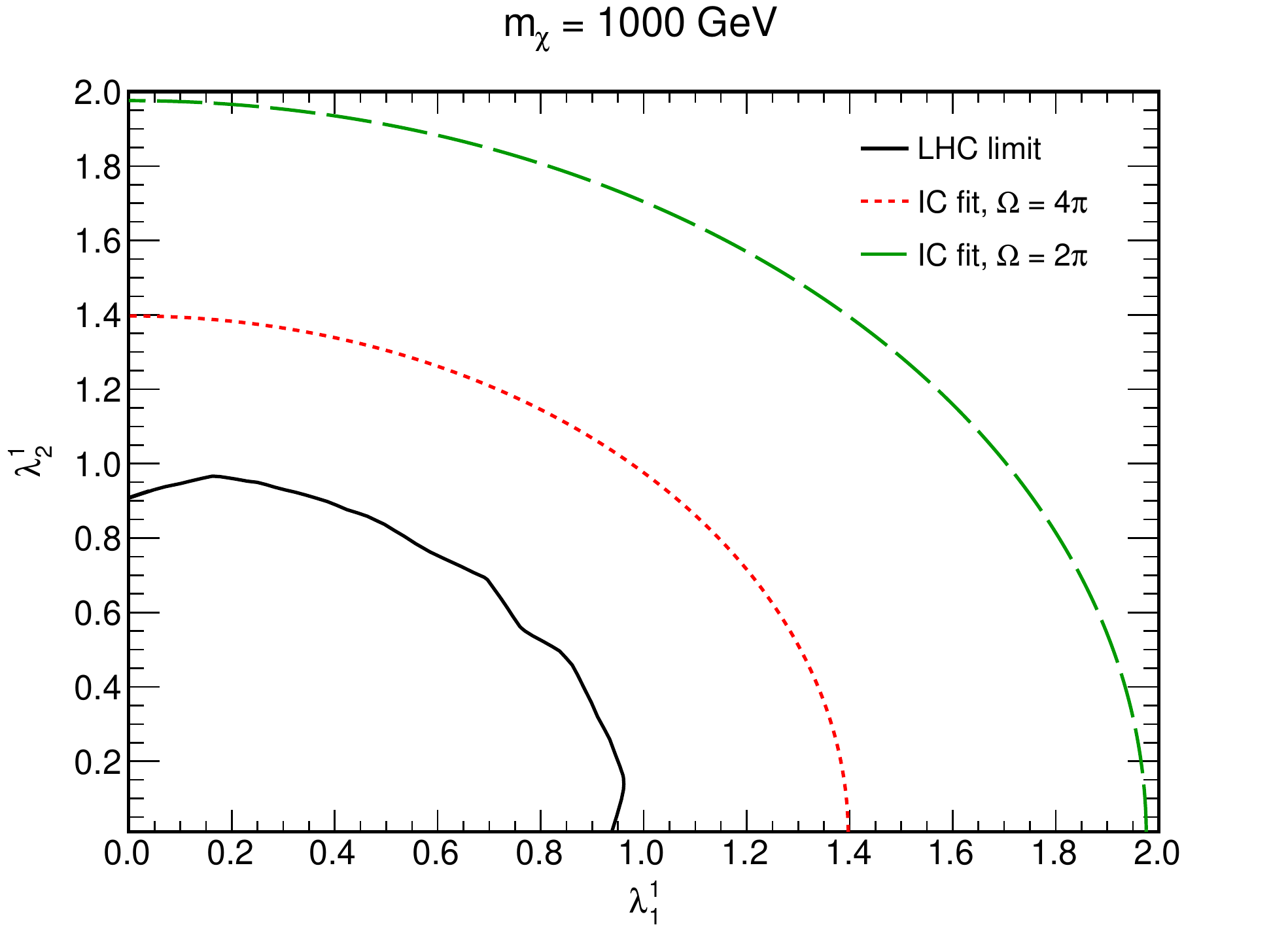}}\\[2mm]
%\hspace*{1.2cm}
\subfloat{\includegraphics[scale=0.36]{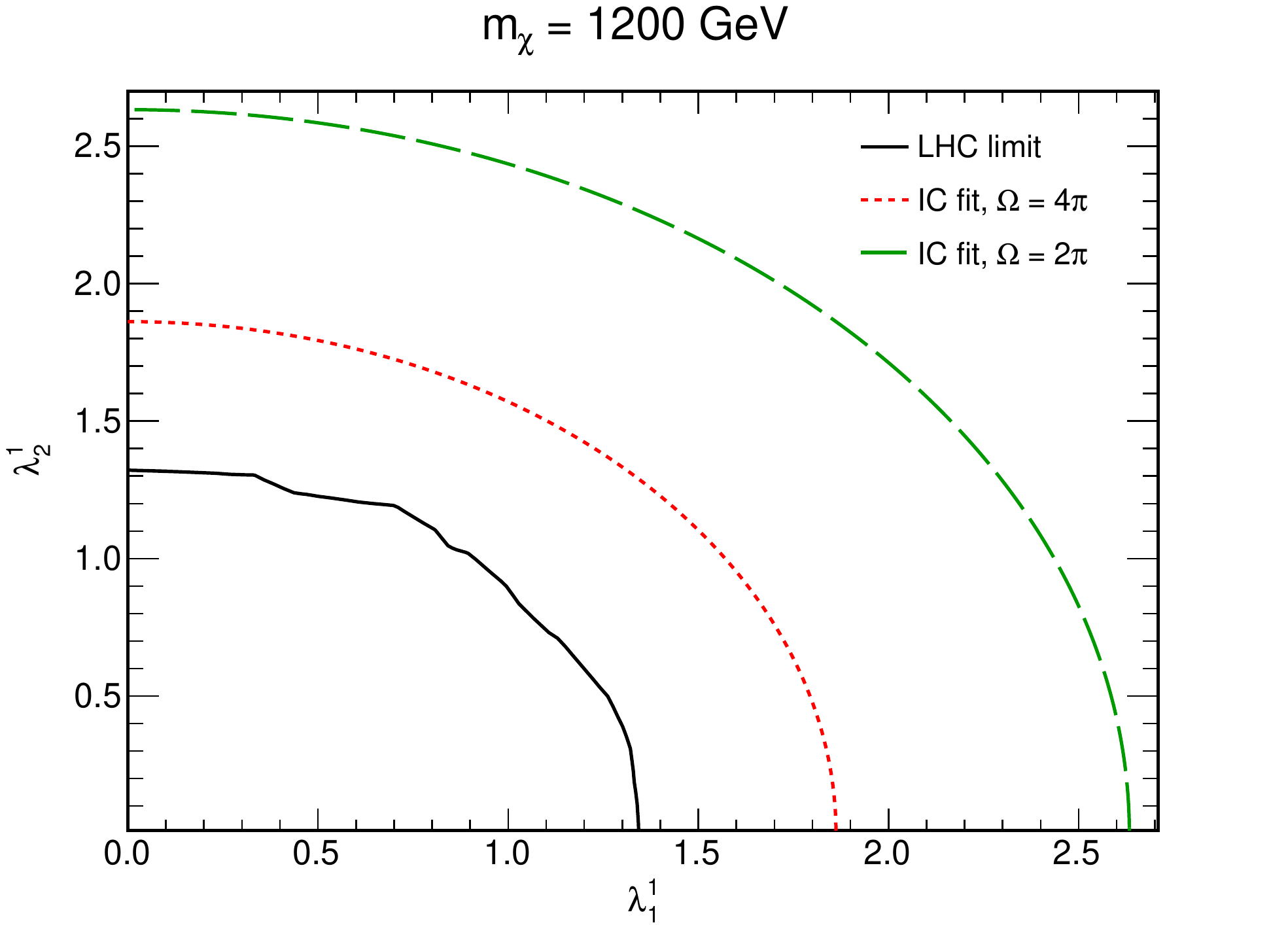}}
\caption{Contours corresponding to the fit of the IceCube spectrum along with the $95\%\,\mathrm{CL}$ LHC upper limits for $m_{\chi}=800\,\mathrm{GeV}$, $1000\,\mathrm{GeV}$ and $1200\,\mathrm{GeV}$.}
\label{fig6.1}
\end{figure}
\end{center}
\vspace*{-4mm}
%%%%%%%%%%%%%%%%%%%%%%%%%%%%%%%%%%%%%%%%%%%%%%%%%%%%%%%%%
The estimates of the parameter $\mu$ for different LQ masses shown in table \ref{table3.1} can be confronted with the constraints arising from the $8\,\mathrm{TeV}$ LHC data. In figure~\ref{fig6.1}, we display the $\hat{\mu}$ values obtained from the fit to the IceCube data along with the $95\%\,\mathrm{CL}$ LHC limits for the masses $800\,\mathrm{GeV}$, $1000\,\mathrm{GeV}$ and $1200\,\mathrm{GeV}$. We see that in the case of a fully opaque Earth, $\Omega = 2\pi\,\mathrm{sr}$, the contours preferred by the IceCube data are excluded by the LHC upper limits at $95\%\,\mathrm{CL}$. This is also the case when no attenuation effects are considered, $\Omega=4\pi\,\mathrm{sr}$, for $1000\,\mathrm{GeV}$ and $1200\,\mathrm{GeV}$, while for $800\,\mathrm{GeV}$ the ranges $0.20<\lambda^1_1<0.40$ and $0.55<\lambda^1_1<0.67$ are not ruled out.\par
After analyzing the low energy constraints on our framework, we are led to conclude that a scenario of interest will include a very suppressed value of $\lambda^{1}_{1}$ compared to $\lambda^{1}_{2}$. This resulted from a combination of the rare $\mu\to e\gamma$ decay and atomic parity violation measurements and our working assumption that $\chi$ coupled primarily the first family of quarks to the electron and the muon. Therefore, by taking $|\lambda^1_1|$ sufficiently small and $|\lambda^1_2|\sim 1.47$, the parameters that give the best fit of the IceCube data, $m_{\chi}=1025\,\mathrm{GeV}$ and $\hat{\mu}=2.4$, are compatible with the low energy constraints and also, as said above, with the dedicated searches of LQs at the LHC. However, as shown in figure~\ref{fig6.1}, this scenario for the LQ triplet is clearly in tension with the $8$ TeV LHC constraints. On the other hand, even though in the idealized case of $\Omega=4\pi\,\mathrm{sr}$ a LQ with mass around $800\,\mathrm{GeV}$ is not ruled out by these constraints, its contribution to the spectrum above PeV is not significant (see figure~\ref{fig3.6}). Furthermore, such a value for the LQ mass is in conflict with the LHC 13 TeV dedicated searches if one requires the LQ to decay only to electrons and muons. Loosing this requirement with an additional decay mode will necessitate a more dedicated recast of LHC searches.

%Furthermore, we observe that after analyzing the low energy constraints on our framework, we are lead to conclude that a scenario of interest will include a very suppressed value of $\lambda^{1}_{1}$ compared to $\lambda^{1}_{2}$. This resulted from a combination of the rare $\mu\to e\gamma$ decay and atomic parity violation measurements and our working assumption that $\chi$ coupled primarily the first family of quarks to the electron and the muon. In light of the $8$ TeV LHC constraints discussed above, a LQ with a mass in the range $800$-$900$ GeV would appear as a viable explanation for the excess of ultra-high energy neutrino events observed at IceCube. However, this mass range is in tension with LHC 13 TeV LQ searches if one requires that our LQ only decays to electrons and muons. Loosing this requirement with an additional decay mode will necessitate a more more dedicated recast of LHC searches.
%
\bigskip
\noindent
{\bf Acknowledgments}
\noindent 
This work has been partially supported by ANPCyT under grants No. PICT
2013-0433 and No. PICT 2013-2266, and by CONICET (NM, AS). ADP was supported in part by the Natural Sciences and Engineering Research Council of Canada (NSERC). In addition, the authors have benefited from conversations with Estefania Coluccio Leskow.
\appendix
\section{Attenuation effects in the Earth}
\label{A1}
The rate of upward-going neutrinos is reduced due to the interactions of the incoming neutrinos with the nucleons in the Earth. Although the interactions with electrons can be  important at $E_{\nu}\simeq 6.3\,\mathrm{PeV}$, where the resonant production of the $W$ boson takes place, we will focus in this appendix on the neutrino-nucleon interactions. The water equivalent interaction length due to neutrino-nucleon interactions is given by
\beq
\label{eqA.1}
L_{\mathrm{int}}=\frac{1}{\sigma_{\nu N}(E_{\nu})N_A},
\eeq 
where $N_A=6.022\times 10^{23}\,\mathrm{cm}^{-3}$ (water equivalent) is Avogadro's number. We note that every neutrino (antineutrino) flavor has a different interaction length according to the specific cross section describing its interactions with nucleons.\par
\begin{center}
\begin{figure}[ht]
\centering
%\hspace*{4mm}
\includegraphics[scale=0.43]{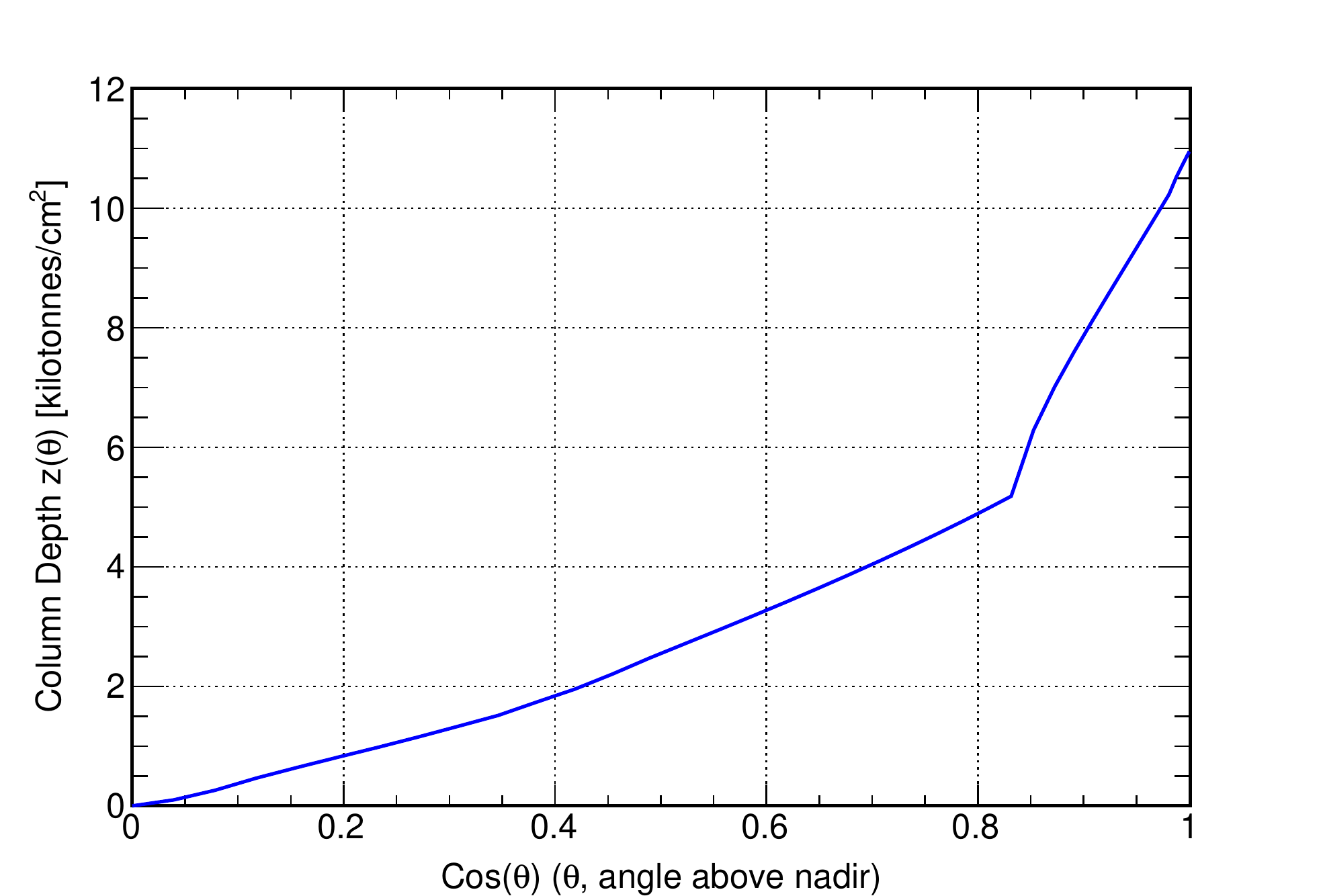}
\centering
\caption{Column depth as a function of the angle of incidence of the incoming neutrinos.}
\label{figA.1}
\end{figure}
\end{center}

 In order to study the attenuation effects, the interaction length should be compared with the ammount of material encountered by an upward-going neutrino, which is konwn as the column depth and depends on the angle of incidence of the incoming neutrinos. The thickness of the Earth as a function of the cosine of the angle of incidence is shown in figure~\ref{figA.1}, where the density profile of the Earth given in ref.~\cite{Gandhi} have been used. The maximum column depth is $11\,\mathrm{kilotonnes/cm^{2}} $, and correponds to a neutrino emerging from the nadir. By plugging the function $z(\theta)$ and the interaction length given in eq.~(\ref{eqA.1}) into the eq.~(\ref{eq3.18}), we obtain the shadow factor $S(E_{\nu})$ (see ref.~\cite{Gandhi}).
%%%%
%%%
\par
As mentioned in sec.~\ref{sec3.2}, the shadow factor depends on the neutrino-nucleon cross section via the interaction length (see eq.~(\ref{eq3.18})) and therefore the addition of the LQ contribution could have in principle an impact on the reduction of the rate of northern events. In order to study this possibility, let us define first the LQ contribution to the total neutrino-nucleon cross section for a flavor $\ell$ as follows
\beq
\label{eqA.2}
\sigma^{\tiny{LQ}}_{\nu_{\ell}N} = |\lambda^1_{\ell}|^2 ~\tilde{\sigma}^{\tiny{LQ}}_{\nu N},
\eeq
where $\lambda^1_{\ell}=\lambda^1_{1,2}$ for $\ell=e,\mu$ respectively, and $\tilde{\sigma}^{\tiny{LQ}}_{\nu N}$ is the same for the two flavors. A similar relation can be written as well for antineutrinos. Also, we denote the SM contribution as $\sigma^{SM}_{\nu N}$ and define
\beq
\label{eqA.3}
L^{\mathrm{tot},(\ell)}_{\mathrm{int}}=\frac{1}{(\sigma^{SM}_{\nu N}+|\lambda^1_{\ell}|^2 ~\tilde{\sigma}^{\tiny{LQ}}_{\nu N})N_A},
\eeq
\beq
\label{eqA.4}
L^{\mathrm{SM},(\ell)}_{\mathrm{int}}=\frac{1}{\sigma^{SM}_{\nu N}N_A}.
\eeq
With these definitions the number of southern events for a given flavor $\ell$ can be written as
\beq
\label{eqA.5}
N^{(\ell)}_{\mathrm{south}}=2\pi\cdot T\cdot N_{\mathrm{eff}}\int \frac{d\phi_{\nu}}{dE_{\nu}}(\sigma^{\mathrm{SM}}_{\nu N}+|\lambda^1_{\ell}|^2 ~\tilde{\sigma}^{\tiny{LQ}}_{\nu N})\,dE_{\nu},
\eeq
and adding all flavors of neutrinos and antineutrinos we obtain
\beq
\label{eqA.6}
N_{\mathrm{south}}=2\pi\cdot T\cdot N_{\mathrm{eff}}\left(3
\int \frac{d\phi_{\nu}}{dE_{\nu}}(\sigma^{\mathrm{SM}}_{\nu N}+\sigma^{\mathrm{SM}}_{\bar{\nu}N})\,dE_{\nu} +
\int \frac{d\phi_{\nu}}{dE_{\nu}}\,\mu (\tilde{\sigma}^{\tiny{LQ}}_{\nu N}+\tilde{\sigma}^{\tiny{LQ}}_{\bar{\nu}N})\,dE_{\nu} \right),
\eeq
where $\mu = |\lambda^1_1|^2+|\lambda^1_2|^2$. On the other hand, the northern events are written as
\beq
\label{eqA.7}
N_{\mathrm{north}}=N^{(\tau + \bar{\tau})}_{\mathrm{north}}+N^{(e + \bar{e})}_{\mathrm{north}}+N^{(\mu + \bar{\mu})}_{\mathrm{north}},
\eeq
with
\bea
N^{(\tau + \bar{\tau})}_{\mathrm{north}} &=& 2\pi \cdot T\cdot N_{\mathrm{eff}}\! \int \frac{d\phi_{\nu}}{dE_{\nu}}\left(S^{\mathrm{SM}}_{\nu}\,\sigma^{\mathrm{SM}}_{\nu N}+S^{\mathrm{SM}}_{\bar{\nu}}\,\sigma^{\mathrm{SM}}_{\bar{\nu}N}\right)\, dE_{\nu}\,, \\
N^{(e + \bar{e})}_{\mathrm{north}} &=& 2\pi \cdot T\cdot N_{\mathrm{eff}}\! \int \frac{d\phi_{\nu}}{dE_{\nu}}\left\{S^{\mathrm{total}}_{\nu_{e}}\left(\sigma^{\mathrm{SM}}_{\nu N}+|\lambda^1_1|^2\tilde{\sigma}^{\tiny{LQ}}_{\nu N}\right)+S^{\mathrm{total}}_{\bar{\nu}_e}\left(\sigma^{\mathrm{SM}}_{\bar{\nu}N}+|\lambda^1_1|^2\tilde{\sigma}^{\tiny{LQ}}_{\bar{\nu}N}\right)\right\}\, dE_{\nu}\,,  \\
N^{(\mu + \bar{\mu})}_{\mathrm{north}} &=&  2\pi \cdot T\cdot N_{\mathrm{eff}}\! \int \frac{d\phi_{\nu}}{dE_{\nu}}\left\{S^{\mathrm{total}}_{\nu_{\mu}}\left(\sigma^{\mathrm{SM}}_{\nu N}+|\lambda^1_2|^2\tilde{\sigma}^{\tiny{LQ}}_{\nu N}\right)+S^{\mathrm{total}}_{\bar{\nu}_{\mu}}\left(\sigma^{\mathrm{SM}}_{\bar{\nu}N}+|\lambda^1_2|^2\tilde{\sigma}^{\tiny{LQ}}_{\bar{\nu}N}\right)\right\}\, dE_{\nu}\,,
\eea
where $S^{\mathrm{total}}_{\nu_{\ell}}$ and $S^{\mathrm{SM}}_{\nu}$ are obtained from eq.~(\ref{eq3.18}) by using the interaction lengths in eqs.~(\ref{eqA.3}) and (\ref{eqA.4}), respectively.\par
\begin{center}
\begin{figure}[ht]
\centering
%\hspace*{-0.4cm}
\hspace*{-2mm}
\subfloat{\includegraphics[scale=0.44]{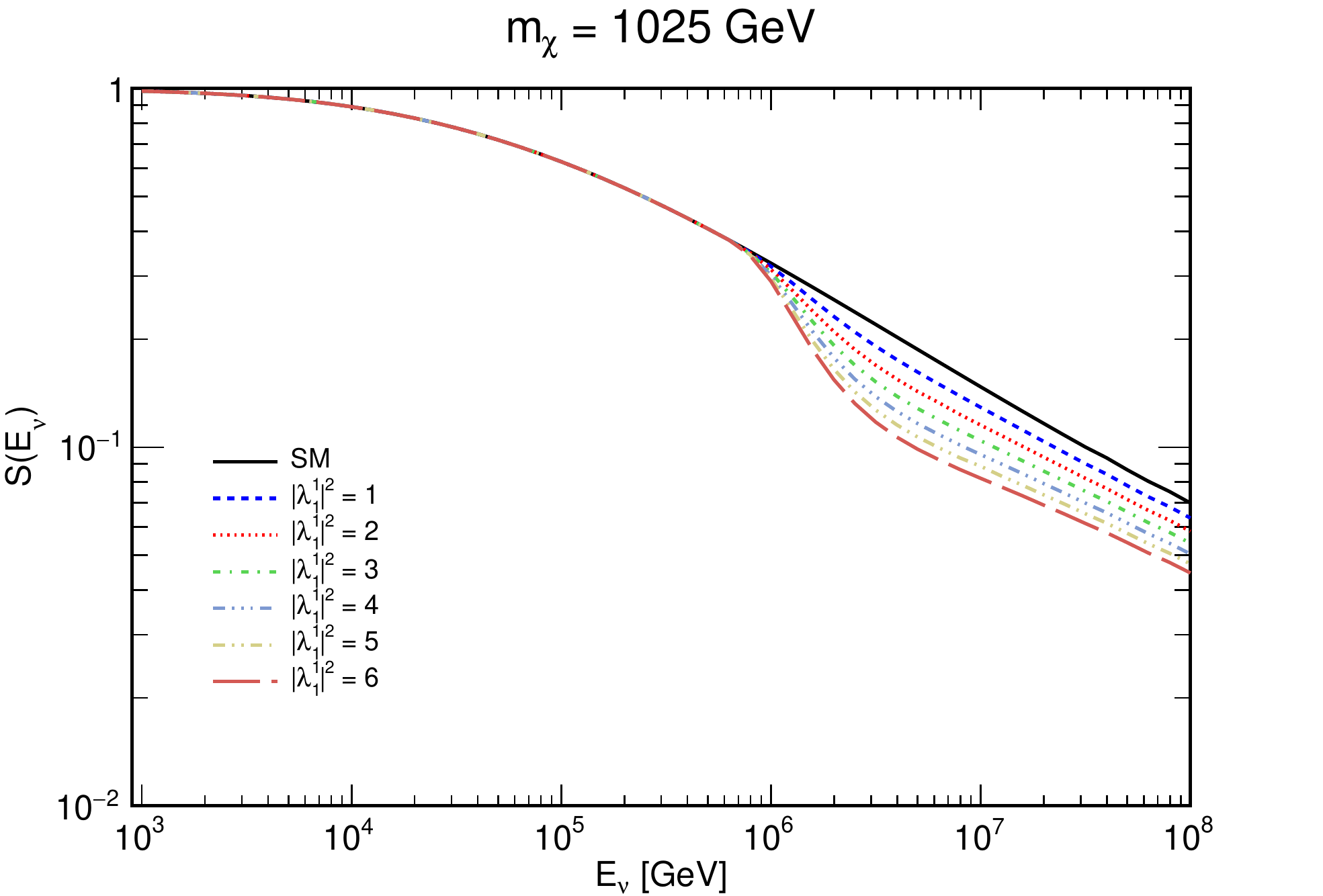}}
%\hspace*{0.2\textwidth}
\hspace*{1mm}
%\label{fig1a}}
\subfloat{\includegraphics[scale=0.44]{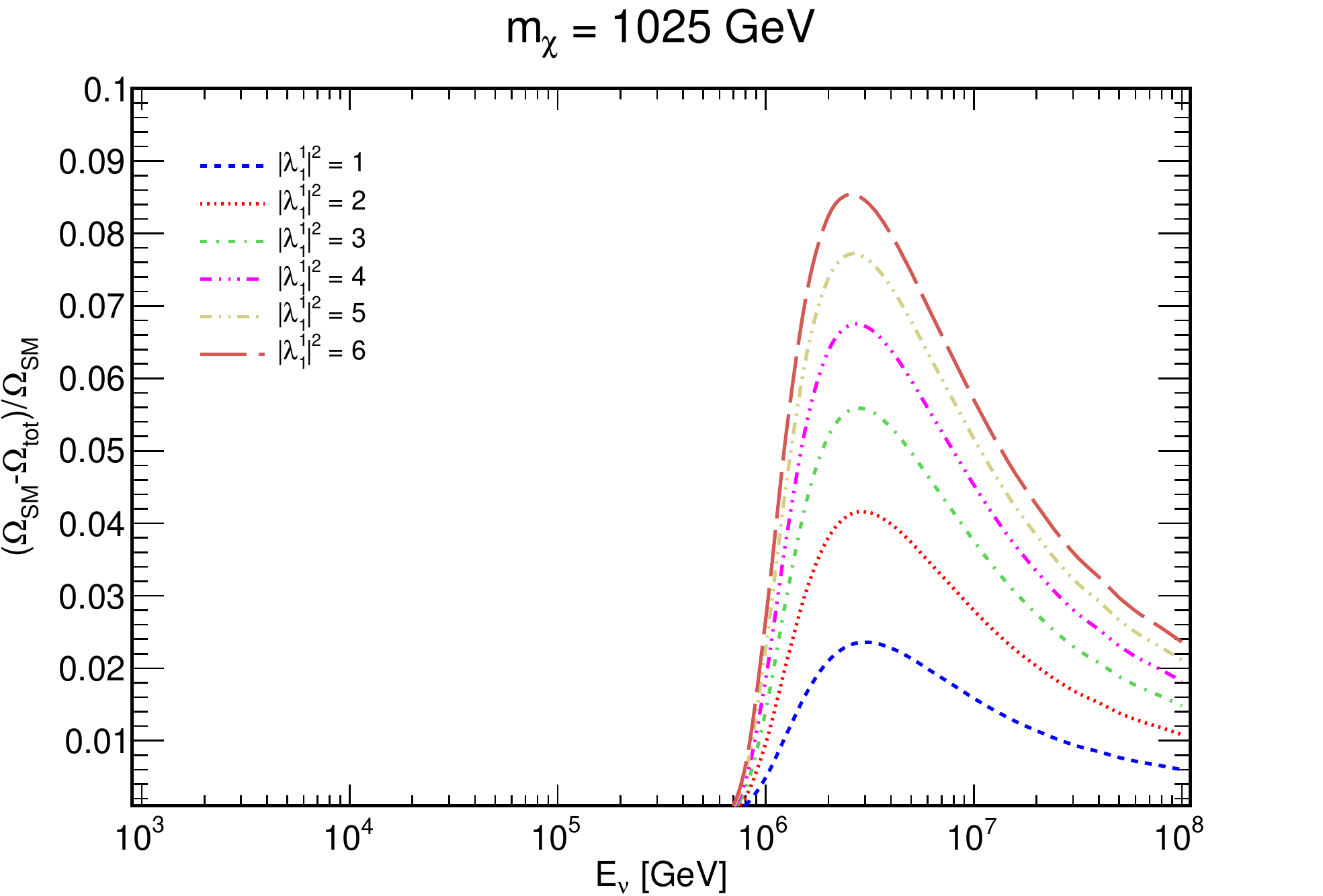}}
\hspace*{4mm}
\caption{Shadow factor corresponding to $\nu_e$ for the SM and adding LQ contributions of various strengths (left) and the respective relative difference between the effective solid angles (right).} 
\label{figA.2}
\end{figure}
\end{center}

In figure~(\ref{figA.2}), we show the $\nu_e$ shadow factor for the SM hypotheses along with the deviations produced by adding different LQ contributions. These contributions correspond to the mass that gives the best fit to the Icecube data, $m_{\chi}= 1025\,\mathrm{GeV}$ (see sec.~\ref{sec3.3}), and $|\lambda^1_1|^2=1$-$6$. Moreover, we display the relative difference between the effective solid angle for the SM hypothesis, $\Omega_{\mathrm{SM}}\equiv 2\pi (1+S^{\mathrm{SM}}_{\nu_e})$, and for the SM+LQ hypothesis, $\Omega_{\mathrm{tot}}\equiv 2\pi(1+S^{\mathrm{total}}_{\nu_{e}})$. From the left plot, we see that the shadow factor is a decreasing function of the incoming neutrino energy that, as expected, begins to deviate from the SM behaviour above the energy threshold associated to the specific LQ contribution, namely $m^2_{\chi}/2M_N$. However, the deviation due to the addition of the LQ contribution is not meaningful as can be concluded from the right plot in figure~\ref{figA.2}. Indeed, the relative difference between the effective solid angles is less than $9\%$, even for a squared coupling as large as $6$. We note that the case of the muonic neutrino is entirely analogous (with the replacement $\lambda^1_1\to\lambda^1_2$).  
%%%
%%%
\par
\begin{center}
\begin{figure}[ht]
\centering
%\hspace*{4mm}
\includegraphics[scale=0.43]{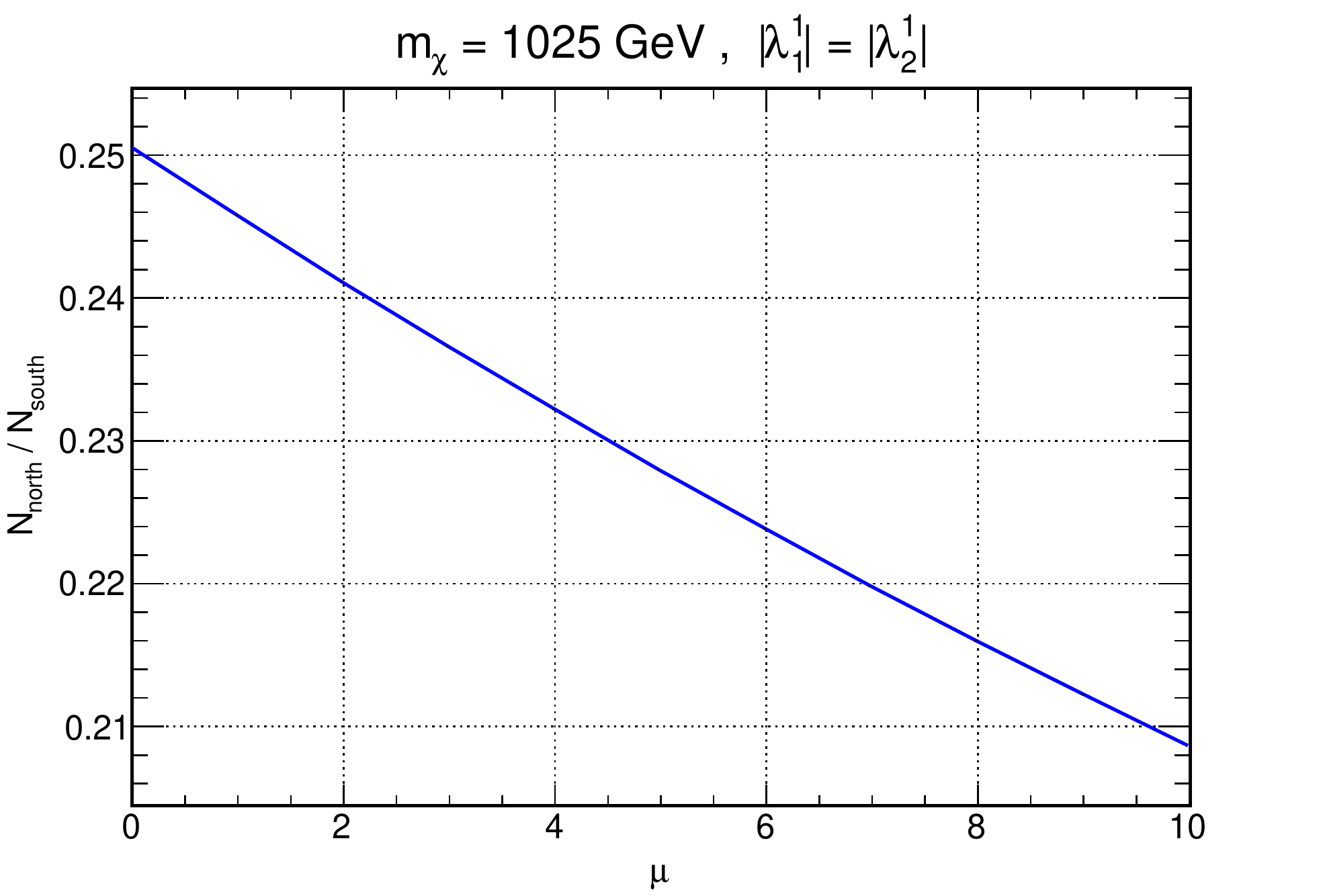}
\caption{Ratio between northern and southern events obtained for the SM ($\mu = 0$) and adding a LQ contribution corresponding to $m_{\chi}=1025\,\mathrm{GeV}$ and different values of $\mu$.}
\label{figA.3}
\end{figure}
\end{center}

Finally, the ratio between southern and northern events, $R\equiv N_{\mathrm{north}}/N_{\mathrm{south}}$, can be computed by using eqs.~(\ref{eqA.6}) and (\ref{eqA.7}). Given a certain value of $\mu$, we can parameterize the couplings as $|\lambda^1_1|=\sqrt{\mu}\cos\alpha$ and $|\lambda^1_2|=\sqrt{\mu}\sin\alpha$. With this choice, $R$ depends on both $\mu$ and the angle $\alpha$. However, by scanning over different values of $\alpha$, we have checked that the variation of $R$ with this angle is very small, with the maximum value of the ratio being obtained for $\alpha=\pi/4$. Thus, we have set $|\lambda^1_1|=|\lambda^1_2|$ and computed the ratio 
$R$ for $\mu$ in the range $0$-$10$. From figure~\ref{figA.3}, we see that the ratio decreases as $\mu$ increases, but the deviation of the SM expectation is at most $17\%$ for $\mu$ as large as $10$. This result is consistent with the conclusions derived from figure~\ref{figA.2}; since the interaction length is dominated by the SM contribution, the impact of the LQ contribution in the disbalance between events coming from the two hemispheres is not significant and this is reflected in the plot of $R$ as a function of $\mu$.
%%%%%%%%%%%%%%%%%%%%%%%%%%%%%%%%%%%%%%%%%%%%%%%%%%%%%%%%%%%%
%%%%%%%%%%%%%%%%%%%%%%%%%%%%%%%%%%%%%%%%%%%%%%%%%%%%%%%%%%%%
\bibliographystyle{JHEP}

\end{document}